\documentclass[aps,pra,twocolumn,showpacs,superscriptaddress,floatfix]{revtex4-1}
\usepackage[colorlinks=true, allcolors=blue]{hyperref}
\usepackage{graphicx,amsmath,amssymb,amsfonts,dsfont,mathtools,dcolumn,bm,physics}
\begin{document}
\preprint{APS/123-QED}
\title{Few-photon transport in Fano-resonance waveguide geometries}
\date{\today}
\author{Kristoffer B. Joanesarson}
\affiliation{Department of Photonics Engineering, DTU Fotonik, Technical University of Denmark, Building 343, 2800 Kongens Lyngby, Denmark}
\author{Jake Iles-Smith}
\affiliation{Department of Photonics Engineering, DTU Fotonik, Technical University of Denmark, Building 343, 2800 Kongens Lyngby, Denmark}
\author{Mikkel Heuck}
\affiliation{Department of Photonics Engineering, DTU Fotonik, Technical University of Denmark, Building 343, 2800 Kongens Lyngby, Denmark}
\author{Jesper M{\o}rk}
\email{jesm@fotonik.dtu.dk}
\affiliation{Department of Photonics Engineering, DTU Fotonik, Technical University of Denmark, Building 343, 2800 Kongens Lyngby, Denmark}
\begin{abstract}
We present a theoretical study of Fano interference effects in few-photon transport. Under appropriate conditions, a local defect in an optical waveguide induces a highly asymmetric transmission lineshape, characteristic of Fano interference. For a two-level emitter placed adjacent to such a defect, here modeled as a partially transmitting element, we find an analytical expression for the full time evolution of single-photon wavepackets and the emitter excitation probability. We show how the partially transmitting element affects the emitter lifetime and shifts the spectral position of the effective system resonances. Using input-output formalism, we determine the single and two-photon $ S $-matrices for both a two-level emitter and a cavity-emitter system coupled to a waveguide with a partially transmitting element. We show how the Fano interference effect can be exploited for the implementation of a Hong-Ou-Mandel switch in analogy with a tunable linear or nonlinear beam splitter.
\end{abstract}
\maketitle
%%%%%%%%%%%%%%%%%%%%%%%%%%%%%%%%%%%%%%%%%%%%%%%%%%%%%%%%%
%%%%%%%%%%%%%%%% INTRODUCTION %%%%%%%%%%%%%%%%%%%%%%%%%%%
\section{\label{sec:intro}Introduction}
A long standing goal for quantum optics is the deterministic control of light at the single-photon level~\cite{kimble2008theQuantum_nature}.  Optical waveguides coupled to (artificial) atoms are examples of promising candidates for guiding and manipulating the photonic qubits with the potential for scalable on-chip quantum information processing~\cite{lodahl2015interfacing_revModPhys}. To efficiently manipulate the optical quantum states one can exploit that strong light-matter coupling can mediate photon-photon interactions and give rise to highly non-linear properties~\cite{chang2014quantum_natPhot,roy2017colloquium_revmodPhys}. Examples of waveguide geometries with demonstrated strong light-matter interactions include nanofibres coupled to atoms~\cite{mitsch2014quantum_natComm}, superconducting waveguides coupled to tunable plasmon qubits~\cite{mirhosseini2019cavity_nature}, plasmonic nanowires~\cite{akimov2007generation_nature}, and photonic crystal waveguides coupled to semiconductor quantum dots~\cite{arcari2014nearUnity_physRevLett,javadi2015single_natComm}.

\begin{figure}[htp]
	\centering
	\includegraphics[width=\columnwidth]
	{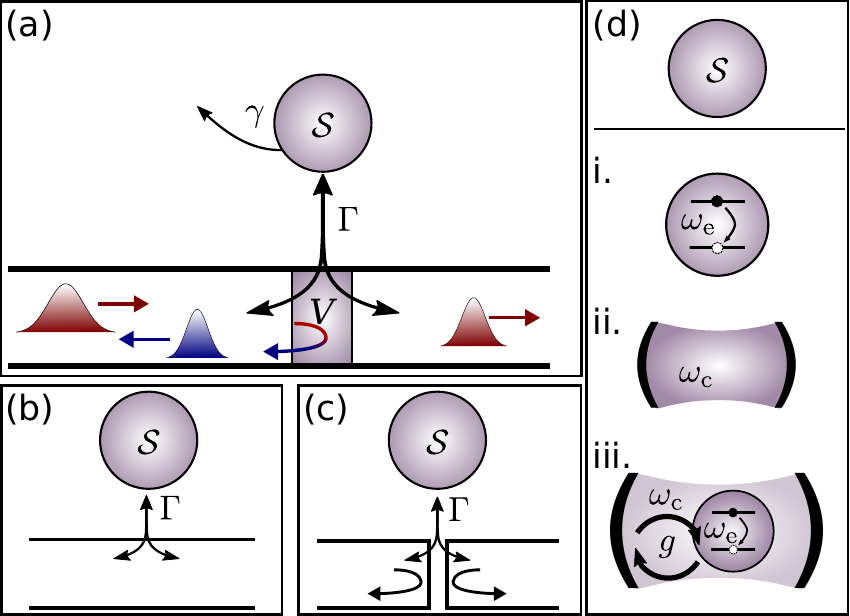}
	\caption{Waveguide geometries. \textbf{(a)} A Fano-resonance waveguide geometry coupled to a quantum system $ \mathcal{S} $ with coupling rate $ \Gamma $ and coupled to non-guided modes with coupling rate $ \gamma $. The waveguide has a PTE in the mirror-symmetry line with strength $ V $. \textbf{(b)} and \textbf{(c)} show the limiting cases of a quantum system side coupled to a waveguide, and coupled to a blocked waveguide, respectively. {\bf{(d)}} The specific quantum systems $ \mathcal{S} $ analyzed in this work. We consider i. a TLE with energy separation $ \hbar\omega_{\mathrm{e}} $, ii. a cavity with resonance frequency $ \omega_{\mathrm{c}} $, and iii. a combined cavity-emitter system with coupling coefficient $ g $.}	\label{fig:sketch_waveguide_PTE}
\end{figure}

Within few-photon scattering experiments and theoretical studies, the waveguide geometry is often modelled as an infinite or semi-infinite collection of harmonic oscillators coupled to one or more discrete-energy-level quantum systems such as a single two-level emitter (TLE)~\cite{shen2007strongly_PRA,nysteen2015scattering_NJoP}, a Jaynes-Cummings (JC) system (i.e. a TLE coupled to a cavity)~\cite{shi2011twoPhoton_PRA,shen2009theory_1_PRA,rephaeli2012fewPhoton_IEEE}, multi-level emitters~\cite{witthaut2010photonNJoP,zheng2013waveguideQED_PRL,das2018photon_1_PRA,das2018photon_2_PRA}, multiple emitters~\cite{rephaeli2011fewPhoton_PRA,cheng2017waveguide}, whispering gallery mode cavities~\cite{shen2009theory_2_PRA}, and a Kerr nonlinear cavity~\cite{liao2010correlated_PRA,xu2014strongly_PRA}. Common for these studies is that the transport properties for the waveguide geometries are almost always assumed to depend only on the specific type of discrete-energy-level quantum system and not on the waveguide properties. In this work we investigate the few-photon transport properties in waveguides having a local defect placed in the mirror-symmetry line of a quantum system~\cite{heuck2013optical_OptLett}, as sketched in Fig.~\ref{fig:sketch_waveguide_PTE}(a). In Fig.~\ref{fig:sketch_waveguide_PTE}(a) an initially right propagating optical pulse is partially transmitted and reflected after interacting with a scattering system, $ S $, and a partially transmitting element (PTE). $ S $ is coupled to the waveguide with a rate $ \Gamma $, the PTE is characterized by a strength $ V $, and dissipation is included through coupling to non-guided modes with leakage rate $ \gamma $. In Fig.~\ref{fig:sketch_waveguide_PTE}(b) and~\ref{fig:sketch_waveguide_PTE}(c) we sketch two limiting cases where the PTE is absent and with maximal strength, corresponding to a side-coupled system and a blocked waveguide, respectively. In this work we will analyse three specific quantum systems, represented by $ \mathcal{S} $ in Fig.~\ref{fig:sketch_waveguide_PTE}(d): i. a TLE, ii. a cavity,  and iii. a coupled emitter-cavity system.

For the waveguide geometry in Fig.~\ref{fig:sketch_waveguide_PTE}(a), with either of the three aforementioned quantum systems, an incident photon far detuned from the system resonance(s) will only interact with the PTE, which we assume has a flat spectral response, i.e. the transmission and reflection coefficients are frequency independent. In the two limiting cases for the PTE strength, i.e. the waveguide geometries in Fig.~\ref{fig:sketch_waveguide_PTE}(b) and (c), the photons will therefore completely transmit and reflect, respectively. On the other hand,  for an incident photon close to being resonant with the quantum system, interaction-mediated transport can result in (partial) reflection for the situation in Fig.~\ref{fig:sketch_waveguide_PTE}(b), and (partial) transmission for the situation in Fig.~\ref{fig:sketch_waveguide_PTE}(c)~\cite{shen2009theory_1_PRA}. In classical optics, experiments demonstrate that engineering the direct waveguide path, e.g. with a local hole impurity in a photonic crystal waveguide (similar to that depicted in Fig.~\ref{fig:sketch_waveguide_PTE}(a)), can be used to exploit linear and nonlinear scattering effects more efficiently, with potential applications for switching, sensing, and pulse shaping~ \cite{heuck2013optical_OptLett,yu2014fano_applPhysLett,bekele2019inPlane_LPR}. This improvement is due to Fano interference, which is characterized by a highly asymmetric transmission lineshape originating from combining a discrete mode system with a continuum of modes~\cite{suh2003displacementSensitive_applPhysLett,fan2003temporal_JOptSocAmA,limonov2017fano_natPhot}. Here, we analyze the quantum optical scattering properties of Fano-resonance waveguide geometries, from now on referred to as Fano-waveguide geometries, and determine to what extent two-photon nonlinear effects can be enhanced.

The Fano interference effect for one and two-photon scattering in a waveguide coupled to a TLE was studied in the recent work by Xu and Fan~\cite{xu2016fano_PRA}. We extend this analysis of Fano-resonance waveguide geometries in the following ways: (i) We determine analytical expressions for the dynamical evolution of single-photon wavepackets in a waveguide with a PTE coupled to a TLE. (ii) We determine the single and two-photon $ S $-matrix for a JC system coupled to a waveguide with a PTE. (iii) We investigate two-photon scattering for envelopes with finite spectral widths and how the Fano interference affects few-photon switching.

The outline of this paper is as follows. In Sec.~\ref{sec:model}, we introduce the waveguide Hamiltonian, emphasizing the incorporation of the partially transmitting element. In Sec.~\ref{sec:singleExcitationDynamics_TLE} we solve the dynamical single-excitation transport properties for the system sketched in Fig.~\ref{fig:sketch_waveguide_PTE}(a), where $ S $ represents a TLE. In Sec.~\ref{sec:S-matrix} we determine the single and two-photon scattering matrices for both a TLE and a cavity-emitter coupled Fano-resonance waveguide geometry. We account here for dissipation via coupling to non-guided modes. In Sec.~\ref{sec:switch} we use our scattering results to analyze few-photon switching properties. Specifically, we investigate a tunable Houng-Ou-Mandel interferometer and how the Fano interference effect may be exploited. Finally, in Sec.~\ref{sec:conclusion}, we summarize our findings.
%%%%%%%%%%%%%%%%%%%%%%%%%%%%%%%%%%%%%%%%%%%%%%%%%%%%%%%%%%
%%%%%%%%%%%%%%%%%%%%% MODEL %%%%%%%%%%%%%%%%%%%%%%%%%%%%%%
\section{\label{sec:model}Waveguide with a partially transmitting element}
The temporal evolution of photons in the waveguide geometry sketched in Fig.~\ref{fig:sketch_waveguide_PTE}(b), in the absence of coupling to a system, is described by the Hamiltonian~\cite{xu2016fano_PRA,fan2010inputoutput_PRA}
\begin{align}
	\begin{split}
		H_{\mathrm{wg},0} &=  \sum_{\mu=R,L}\int_{-\infty}^{\infty} \dd{k}\, v_g \hbar k  \, b_{\mu}^{\dagger}(k)b_{\mu}(k),
	\end{split} \label{eq:H_wg}
\end{align}
where $ v_g $ is the group velocity and $ b_{\mu}^{\dagger} $($ b_{\mu} $) is the bosonic creation (annihilation) operator satisfying the commutator relation $ \comm{b_{\mu}(k)}{b_{\nu}^{\dagger}(k')} = \delta_{\mu\nu}\delta(k-k') $. The subscript $ \mu $ refers to independent waveguide channels corresponding to right (R) and left (L) propagating modes (and later dissipative modes). We have assumed operation in a regime, where a linear approximation to the dispersion relation is valid, and chosen a reference frame where the wavenumbers are centred around zero. From here on, the absence of integration limits indicate integration from $ -\infty $ to $ \infty $. Furthermore, we will work in normalized units with $ \hbar=v_g=1 $, such that energies, wavenumbers, and frequencies are measured in the same units.

To describe the propagation of photons in the waveguide geometry sketched in Fig.~\ref{fig:sketch_waveguide_PTE}(a) with $ \Gamma=0 $, we add to the Hamiltonian in Eq.~\eqref{eq:H_wg} a term corresponding to a local change in the optical density of states. We model the combined waveguide and PTE using the Hamiltonian $ H_{\mathrm{wg}} = H_{\mathrm{wg},0} + H_{\mathrm{PTE}} $, with
\begin{align}
	H_{\mathrm{PTE}} = V\int \frac{\dd{k}}{\sqrt{2\pi}}\int \frac{\dd{ k'}}{\sqrt{2\pi}}b_R^{\dagger}(k)b_L(k') + \mathrm{h.c.}, \label{eq:H_PTE}
\end{align}
as used in~\cite{xu2016fano_PRA}. Using input-output formalism~\cite{collett1984squeezing_PRA,gardiner1985input_PRA,gardiner1987input_JOptSocAmB} we determine the scattering properties for the waveguide with a PTE. We start by defining the single-photon scattering eigenstates~\cite{fan2010inputoutput_PRA}
\begin{align}
	\ket{k^+}_{\nu} = b^{\dagger}_{\mathrm{in},\nu}(k)\ket{\varnothing}, \qquad \ket{p^-}_{\mu} = b^{\dagger}_{\mathrm{out},\mu}(p)\ket{\varnothing},
\end{align}
where $  |\varnothing \rangle $ is the vacuum state and $ b^{\dagger}_{\mathrm{in(out)},\nu}(k) $ is the Fourier transformed input (output) field operator for the mode $ \nu $ defined as
\begin{align}
	b_{\text{in(out)},\nu}(t) &= \int \frac{\dd{k}}{\sqrt{2\pi}}\, b_{\nu}(k,t_{\mathrm{in(out)}})e^{ik\qty(t_{\mathrm{in(out)}}-t)}, \label{eq:inOut:inPut-gen_index_a}
\end{align}
where $ b_{\nu}(k,t)=e^{iHt}b_{\nu}(k)e^{-iHt} $ is in the Heisenberg picture. Photons are incident in the waveguide at large negative times, $ t_{\mathrm{in}}\rightarrow -\infty $, and exit the waveguide at $ t_{\mathrm{out}}\rightarrow \infty $.  Assuming $ V\in \mathbb{R} $, the input-output relation in vector form is
\begin{align}
	\mathbf{b}_{\text{out}}(t) &= \mathbf{C}\mathbf{b}_{\text{in}}(t), 
\end{align}
with
\begin{align}
	\mathbf{C} &=\mqty(t_B & r_B \\ r_B & t_B) =\frac{1}{1+\qty(\frac{V}{2})^2}\mqty( 1-\qty(\frac{V}{2})^2 & -iV
	\\
	-iV & 1-\qty(\frac{V}{2})^2 ),
\end{align}
as shown by Xu and Fan~\cite{xu2016fano_PRA}. We interpret the dimensionless parameter $ V $ as the PTE strength, as it determines the proportion of reflected and transmitted light characterized by the frequency independent transmission and reflection coefficients $ |t_B|^2 $ and $ |r_B|^2 $, respectively. We note that $ |t_B|^2 + |r_B|^2 = 1 $ as required by the loss-less PTE. Furthermore, we note that both coefficients are uniquely confined in the interval $ [0, \, 1] $ for $ V\in [0,\, 2] $. $ V=0 $ and $ V=2 $ corresponds to having a waveguide geometry like that of Fig.~\ref{fig:sketch_waveguide_PTE}(b) and that of Fig.~\ref{fig:sketch_waveguide_PTE}(c), respectively. $ V=2/\qty(1+\sqrt{2}) $ gives a balanced PTE i.e. $ |t_B|^2 = |r_B|^2 = 1/2 $.
%%%%%%%%%%%%%%%%%%%%%%%%%%%%%%%%% SINGLE EXCITATION DYNAMICS %%%%%%%%
%%%%%%%%%%%%%%%%%%%%%%%%%%%%%%%%%%%%%%%%%%%%%%%%%%%%%%%%%%%%%%%%%%%%%
\section{A Fano-resonance waveguide system with a two-level emitter -- exact dynamics}\label{sec:singleExcitationDynamics_TLE}
In this section we consider a TLE symmetrically and frequency-independently coupled to a linearly dispersive waveguide with a PTE in the mirror-symmetry line. This particular geometry corresponds to the illustration in Fig.~\ref{fig:sketch_waveguide_PTE}(a) with the TLE in place of $ \mathcal{S} $ (System i. in Fig.~\ref{fig:sketch_waveguide_PTE}(d)). The Hamiltonian is
\begin{align}
	\begin{split}
		H &= \omega_{\mathrm{e}}\sigma_{+}\sigma_{-} + \sum_{\mu = L,R} \int \dd{k}\, k b_{\mu}^{\dagger}(k)b_{\mu}(k)
		\\
		& \quad +  V \int \frac{\dd{k}}{\sqrt{2\pi}}\frac{\dd{k'}}{\sqrt{2\pi}} \, \qty[ b_R^{\dagger}(k)b_L(k') + \mathrm{h.c.} ]
		\\
		& \quad  + \int \dd{k} \frac{\sqrt{\Gamma/2}}{\sqrt{2\pi}}\qty(\sigma_{-} \sum_{\mu=R,L} b_{\mu}^{\dagger}(k) + \mathrm{h.c.}).
	\end{split} \label{eq:Hprime_a}
\end{align}
The first term describes the bare waveguide with a PTE, as introduced in Sec.~\ref{sec:model}. The second term governs the free evolution of the TLE, where $ \sigma_{+} $($ \sigma_{-}$) is the raising (lowering) operator.  The last term accounts for the emitter-waveguide coupling in the dipole and rotating wave approximation~\cite{scully_zubairy_1997}. Because $ [H,\hat{N}]  = 0$, where $ \hat{N} = \sigma_{+}\sigma_{-} + \sum_{\mu=L,R}\int \dd{k}\, b_{\mu}^{\dagger}(k)b_{\mu}(k) $ is the excitation number operator, we can solve the dynamics in the single-excitation manifold by considering the single-excitation state vector expansion
\begin{align}
	\ket{\psi(t)} = \chi(t)\sigma_{+}\ket{\varnothing} + \sum_{\mu=L,R} \int \dd{k} \xi_{\mu}(t,k)b_{\mu}^{\dagger}(k)\ket{\varnothing}.
\end{align}
The coefficient $ \chi(t) $ describes the dynamical emitter excitation with $ |\chi(t)|^2 $ being the probability of finding the TLE in its excited state. The complex envelope shape at time $ t $ for the part of the photon traveling to the right(left) is given by $ \xi_{R(L)}(t,k) $. From the Schr{\"o}dinger equation, we find a set of coupled differential equations for the state vector coefficients
\begin{subequations}
	\begin{align}
		i \dot{\chi}(t) &= \omega_e \chi(t)  +  \frac{\sqrt{\Gamma/2}}{\sqrt{2\pi}}\int \dd{k}\,  \qty[\xi_R(t,k) + \xi_L(t,k)],
		\\
		i\dot{\xi}_R(t,k) &= k\xi_R(t,k)   +  \frac{\sqrt{\Gamma/2}}{\sqrt{2\pi}}\chi(t) + \frac{V}{2\pi}\int \dd{k'}\, \xi_L(t,k'), \label{eq:xi_R_00}
		\\
		i \dot{\xi}_L(t,k) &= k\xi_L(t,k)   +  \frac{\sqrt{\Gamma/2}}{\sqrt{2\pi}}\chi(t) + \frac{V}{2\pi}\int \dd{k'}\, \xi_R(t,k'). \label{eq:xi_L_00}
	\end{align}
\end{subequations}
Without a PTE, i.e. $ V=0 $, the solution is well-known~\cite{chen2011coherent_NJoP,liao2015singlePhoton_PRA}. For a non-zero $ V $ and with known initial conditions, $ \chi(t_i) $, $ \xi_R(t_i,k) $, and $ \xi_L(t_i,k) $, an analytical solution is also possible and given in its exact form in App.~\ref{app:1excDyn_TLE}. In the case $ V\neq 0 $, the emitter excitation coefficient is
\begin{widetext}
	\begin{align}
		\chi(t) &= \chi(t_i)e^{-\qty[\frac{\tilde{\Gamma}}{2}+i\omega_{\mathrm{e}}]\qty(t-t_i)} - i\frac{\tilde{\Gamma}}{2\sqrt{\pi \Gamma}}
		\int \dd{k'}\, \xi(t_i,k')e^{-ik'\qty(t-t_i)}\frac{1-e^{-\qty[\frac{\tilde{\Gamma}}{2}+i\qty(\omega_{\mathrm{e}}-k')]\qty(t-t_i)}}{\frac{\tilde{\Gamma}}{2}+i\qty(\omega_{\mathrm{e}}-k')}, \label{eq:chi_t}
	\end{align}
\end{widetext}
with $ \xi(t_i,k') = \xi_R(t_i,k') + \xi_L(t_i,k') $. We have also defined here an effective complex coupling rate which depends on the PTE stength:
\begin{align}
	\tilde{\Gamma} = \frac{\Gamma}{1+\frac{iV}{2}}. \label{eq:Gamma_tilde}
\end{align}

\subsection{Single-photon emission by an emitter relaxation process}\label{secIII_subA}
Here we study the emitter relaxation process for an initially excited TLE in a waveguide with a PTE, corresponding to the geometry sketched in Fig.~\ref{fig:sketch_waveguide_PTE}(a) with system i. in place of $ \mathcal{S} $. With $ \chi(t_i)=1 $ and $ \xi_R(t_i,k)=\xi_L(t_i,k) = 0 $, Eq.~\eqref{eq:chi_t} reduces to
\begin{align}
	\chi(t) = e^{-\qty[\frac{\tilde{\Gamma}}{2}+i\omega_{\mathrm{e}}]\qty(t-t_i)}.
\end{align}
The effect of the PTE is to increase the emitter lifetime by a factor of $\mathrm{Re}\{\qty(1+iV/2)^{-1}\}^{-1}= 1+(V/2)^2 $. In the long-time limit, $ t_f \longrightarrow \infty $, we find for the photon envelopes
\begin{align}
	\qty|\xi_R(t_{f} , k)|^2 = \qty|\xi_L(t_f , k)|^2= \frac{\frac{\Gamma}{4\pi}}{1+\qty(\frac{V}{2})^2} \qty|\frac{1}{\qty(k-\omega_{\mathrm{e}})+i\frac{\tilde{\Gamma}}{2}}|^2.
\end{align}
A photon generated by the relaxation process will thus have a Lorentzian spectrum even if we introduce a PTE. We see, however, that the center frequency is shifted by $ \mathrm{Im}\{\tilde{\Gamma}/2\}=-\Gamma\frac{V}{2}(1+(V/2)^2)^{-1} $. Both the frequency shift and altered emitter lifetime are characteristic of a Lamb shift and the Purcell effect induced by a change in the local density of states~\cite{rybin2016purcell_scientificReports,lodahl2015interfacing_revModPhys}, here caused by the PTE in the waveguide.

\subsection{Single-photon scattering dynamics}
We can use the analytical expressions for the state vector coefficients to calculate how a single photon scatters off the TLE as well as the probability of finding the emitter in the excited state as a function of time. In the following, we take the input state to be an initially right-propagating single photon, entering from the left side of the waveguide, with a Gaussian spectral envelope 
\begin{align}
	\xi_R(k,t_i) = \qty(\frac{1}{2\pi \sigma^2})^{1/4} e^{-\qty(\frac{k-k_c}{2\sigma})^2}e^{-i t_i \qty(k-k_c)}.\label{eq:Gaussian}
\end{align}
The spectral width, $ \sigma $, and the center frequency $ k_c $ uniquely characterize the spectral shape. We assume that the initial time, $ t_i $, is chosen such that the real-space envelope is far away from the emitter.

\begin{figure*}[htp]
	\centering
	\includegraphics[width=\textwidth]
	{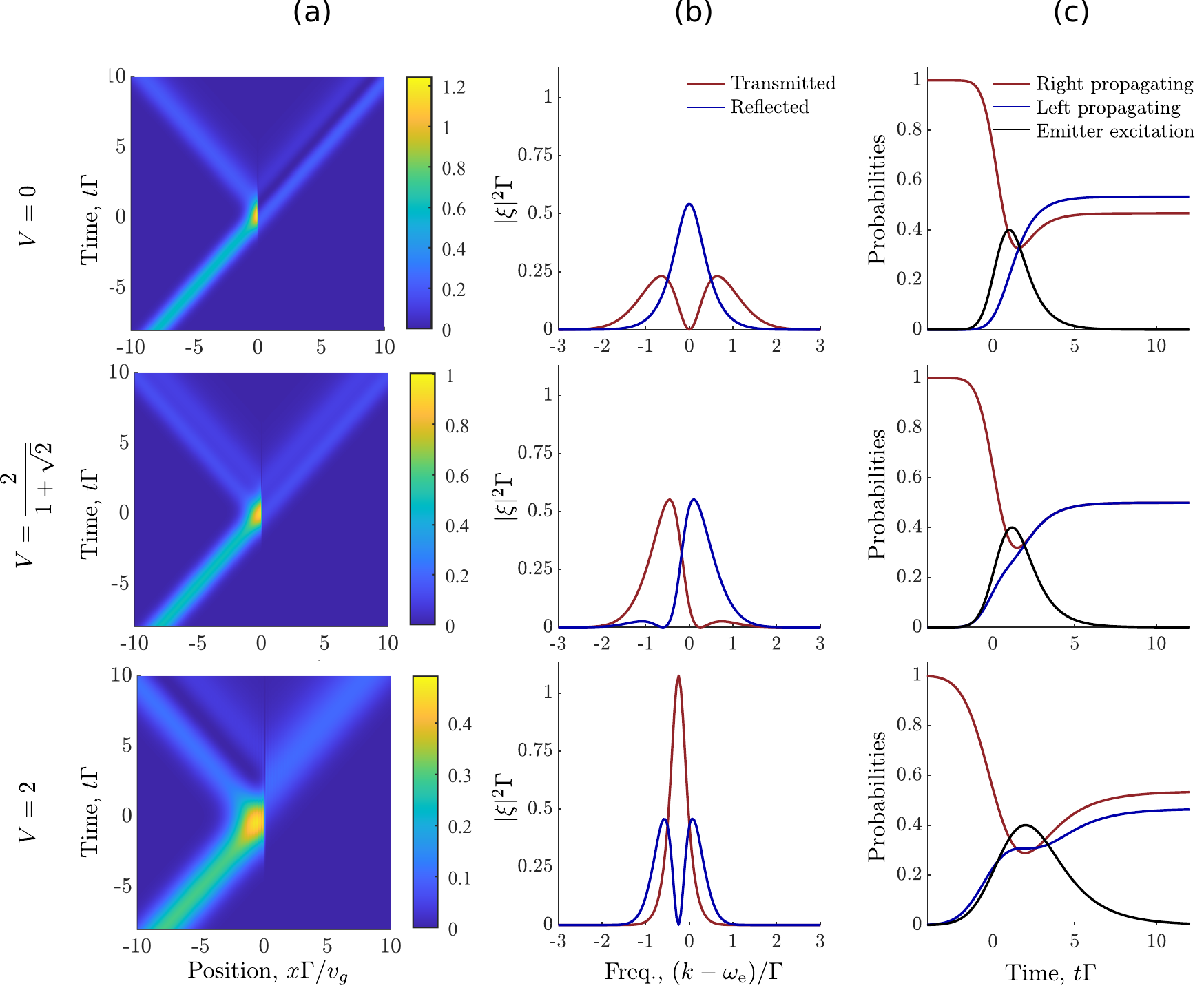}
	\caption{Single-photon scattering in a PTE-waveguide system coupled to a TLE. The input state is an initially right-propagating single photon, entering from the left side of the waveguide, with a Gaussian spectral envelope with $ k_c = \omega_{\mathrm{e}} + \Gamma \times \mathrm{Im}\{\qty(1+iV/2)^{-1}\}/2
		$ and $
		\sigma = \sigma_0 \times  \mathrm{Re}\{\qty(1+iV/2)^{-1}\} $. Column \textbf{(a)}: Single-photon propagating in the waveguide with $ V=0 $ (row 1), $ V=2/\qty(1+\sqrt{2}) $ (row 2), and $ V=2 $ (row 3). The color scale shows the field amplitude vs. position and time. Column \textbf{(b)}: Spectral envelope in the asymptotic limit for the transmitted (red) and reflected (blue) part of the wavepacket.  Column \textbf{(c)}: Emitter excitation probability $ |\chi|^2 $ (black) as well as the probabilities for finding a photon propagating in the right (red) and left (blue) direction versus time.}	\label{fig:dynamical_wavepacket}
\end{figure*}

In the absence of a PTE ($ V=0 $), the emitter excitation probability, $|\chi(t)|^2 $, can reach a maximum of $ 0.4 $ when the envelope is centered around $ \omega_{\mathrm{e}} $ and has a width of $  \sigma_0 \approx 0.73 \Gamma$~\cite{chen2011coherent_NJoP}. If a PTE is introduced, i.e. $ V>0 $, this maximum value is reached by changing the envelope parameters according to
\begin{align}
	k_c &= \omega_{\mathrm{e}} + \frac{\Gamma}{2} \times \mathrm{Im}\qty{\frac{1}{1+\frac{iV}{2}}}\equiv \tilde{\omega}_{\mathrm{e}},
	\\
	\sigma &= \sigma_0 \times  \mathrm{Re}\qty{\frac{1}{1+\frac{iV}{2}}}.
\end{align}
This translation and rescaling, respectively, is again due to a change in the local density of states from introducing a PTE as discussed in Sec.~\ref{secIII_subA}, and is easily verified by insertion into Eq.~\eqref{eq:chi_t}. From here on, we will refer to $ \tilde{\omega}_{\mathrm{e}} $ as the \textit{effective} emitter resonance frequency. In Fig.~\ref{fig:dynamical_wavepacket}, we plot the envelope dynamics as well as the emitter excitation probability and scattering probabilities for the above given envelope parameters using three different values of $ V $, corresponding to an unblocked waveguide ($ V=0 $), a balanced PTE ($ V=\sqrt{2}/\qty(1+\sqrt{2}) $), and a blocked waveguide ($ V=2 $). The initial condition is $ \chi(t_i)=\xi_L(t_i,k)=0 $, and $ \xi_R(t_i,k) $ is given by Eq.~\eqref{eq:Gaussian}. Fig. \ref{fig:dynamical_wavepacket}(a) shows how the photon propagates in the waveguide. When the wavepacket reaches the emitter and the PTE at $ x=0 $, it partially transmits and reflects. In row 1 of Fig.~\ref{fig:dynamical_wavepacket}(a), frequencies far-off-resonance are transmitted as they do not interact with the emitter. This is also evident from row 1 of Fig.~ \ref{fig:dynamical_wavepacket}(b), which shows the asymptotic behavior of the scattered state. In row 2 of Fig.~ \ref{fig:dynamical_wavepacket}(a) the incident photon is scattered into a left and a right propagating part with equal probability. The scattered state of Fig.~\ref{fig:dynamical_wavepacket}(a) row 3 is qualitatively similar to row 1 of Fig.~\ref{fig:dynamical_wavepacket}(a) but with the key difference that the reflected and transmitted envelope shapes have interchanged. This is of course due to the fact that frequencies far-off-resonance are now reflected, whereas frequencies in the vicinity of $ \tilde{\omega}_{\mathrm{e}} $ are transmitted via the interaction with the emitter. The asymptotic output envelopes as well as the input envelope are plotted in Fig. \ref{fig:dynamical_wavepacket}(b), clearly showing which frequencies are reflected and which are transmitted. For the waveguide without the PTE (row 1) and with a fully blocking PTE (row 3), the output spectrum is symmetric around $ \tilde{\omega}_{\mathrm{e}} $. For the balanced PTE (row 2), the transmitted and reflected wavepackets are anti-symmetric around $ \tilde{\omega}_{\mathrm{e}} $. The asymptotic scattering behavior will be analyzed in depth in Sec.~\ref{sec:S-matrix}. In fig.~\ref{fig:dynamical_wavepacket}(c) we plot the probability of measuring the emitter in its excited state as well as measuring a right or a left propagating photon versus time. We see that for all three PTE strengths a maximum excitation probability of $ 0.4 $ occurs for $ t \approx 1/\Gamma $, when the envelope center is close to the emitter position. In row 2 of Fig.~\ref{fig:dynamical_wavepacket}(c), we see that the asymptotic output photons are equal parts reflected and transmitted. This is in fact always true for the balanced PTE in case of a symmetric envelope shape and a center frequency of $ \tilde{\omega}_{\mathrm{e}}$.
%%%%%%%%%%%%%%%%%%%%%%%%%% S-MATRIX CALCULATIONS %%%%%%%%%%%%%%%%%
%%%%%%%%%%%%%%%%%%%%%%%%%%%%%%%%%%%%%%%%%%%%%%%%%%%%%%%%%%%%%%%%%%
\section{Few-photon scattering matrices}\label{sec:S-matrix}
There are many ways of determining the scattering properties for waveguide systems considered in this work. One approach is by determining the full dynamics using a wavefunction approach, as we did in Sec.~\ref{sec:singleExcitationDynamics_TLE}, and calculating the asymptotic behavior. Such an approach can also be used to determine the evolution of a two-photon wavepacket in a waveguide coupled to a JC system~\cite{stoly}. If only the asymptotic wavepackets are of interest, however, determining the scattering matrix -- or $ S $-matrix -- can offer a computational advantage. The $ S $-matrix can be found in a number of ways by using e.g. the Lippmann-Schwinger formalism~\cite{shen2007strongly_PRA},  path-integral techniques~\cite{shi2015multiphoton_PRA}, Dyson series~\cite{hurst2018analytical_PRA}, or via the input-output relations using the method introduced by Fan et al. in~\cite{fan2010inputoutput_PRA}. In this section we will use the latter as this method is both common and transparent. 

\subsection{Single-photon $ S $-matrix}
The single-photon $ S $-matrix, which describes the scattering process of a single-photon input, is defined by 
\begin{align}
	S_{p;k}^{\mu;\nu} = {}_{\mu}\braket{p^-}{k^+}_{\nu} = \mel{\varnothing}{b_{\mathrm{out},\mu}(p)}{k^+}_{\nu}, \label{eq:S_pk_1ph}
\end{align}
where $ \ket{p^{-}} $ and $ \ket{k^{+}} $ are the scattering eigenstates defined in Sec.~\ref{sec:model}. A single-photon state thus scatters according to
\begin{align}
	\ket{\psi_{\mathrm{in}}^{(1)}} &= \sum_{\nu} \int \dd{k}\, \xi_{\mathrm{in},\nu}(k)b_{\nu}^{\dagger}(k)\ket{\varnothing} \nonumber
	\\
	\stackrel{S}{\longrightarrow} & \,\ket{\psi_{\mathrm{out}}^{(1)}} = \sum_{\mu} \int \dd{p}\, \xi_{\mathrm{out},\mu}(p)b_{\mu}^{\dagger}(p)\ket{\varnothing},
\end{align}
where $ \xi_{\mathrm{in},\nu} $($ \xi_{\mathrm{out},\mu} $) is the spectral single-photon input (output) envelope. From the definition of the $ S $-matrix, the single-photon output envelopes are determined from
\begin{align}
	\xi_{\mathrm{out},\mu}(p) = \sum_{\nu}\int \dd{k}\, S_{p;k}^{\mu;\nu} \xi_{\mathrm{in},\nu}(k).
\end{align}
In order to determine $ S_{p;k}^{\mu;\nu} $ we first choose a specific waveguide geometry and determine a set of input-output relations.

\subsection{Input-output relations}
The relation between incoming and outgoing fields can be described in terms of a set of input-output relations~\cite{gardiner1985input_PRA}. In determining the input-output relations we assume a time-independent Hamiltonian similar to that of Eq.~\eqref{eq:Hprime_a} but for an initially unspecified system and system-waveguide Hamiltonian term. Furthermore, we introduce coupling to additional, unspecified, waveguide channels, which account for dissipation into a reservoir~\cite{rephaeli2013dissipation_photResearch}. The Hamiltonian then reads
\begin{align}
	\begin{split}
		H &= \sum_{\mu}\int \dd{k}\, k b_{\mu}^{\dagger}(k)b_{\mu}(k) + H_{\mathrm{s}} + H_{\mathrm{s-wg}}
		\\
		& \quad + \sum_{\mu\neq \nu} V_{\mu,\nu}\int \frac{\dd{k}}{\sqrt{2\pi}}\int \frac{\dd{k'}}{\sqrt{2\pi}}b_{\mu}^{\dagger}(k)b_{\nu}(k').\label{H_general}
	\end{split}
\end{align}
The first and last terms are similar to the waveguide and PTE contributions in Eq.~\eqref{eq:H_wg} and Eq.~\eqref{eq:H_PTE}, respectively. The second and third term correspond to the relevant system and its coupling to the waveguide, respectively. We use the Heisenberg equation of motion, $ \dot{c}=-i\comm{c}{H} $, where $ c $ is an arbitrary operator.
For the waveguide operators we use that $ c=b_{\mu}(k) $ and $ [b_{\mu},H_{\mathrm{s}}]=0 $:
\begin{align}
	\begin{split}
		\dot{b}_{\mu}(t,k) &= -i \comm{b_{\mu}(t,k)}{H_{\mathrm{s-wg}}}
		\\
		& \quad  -ik b_{\mu}(t,k) -i\sum_{\nu/\mu} V_{\mu,\nu}\int \frac{\dd{k'}}{2\pi} b_{\nu}(t,k').
	\end{split}
\end{align}
This, as well as the input-output field operators as defined in Eq.~\eqref{eq:inOut:inPut-gen_index_a}, is sufficient to determine the full set of input-output relations and in turn the single-photon $ S $-matrix. We illustrate this with two paradigmatic examples; first the case of a TLE and then a JC system coupled to a Fano-waveguide.

\subsection{Single-photon $ S $-matrix -- TLE} \label{sec:sMatrix_TLE}
The single-photon $ S $-matrix for a Fano-waveguide geometry coupled to a TLE in the mirror-symmetry line was first calculated in~\cite{xu2016fano_PRA}. We start by summarizing the results of~\cite{xu2016fano_PRA} before extending the analysis to include both a TLE and a cavity. We use the Hamiltonian
\begin{align}
\begin{split}
H &= \omega_{\mathrm{e}}\sigma_{+}\sigma_{-}  + \sum_{\mu}\int \dd{k}\, k b_{\mu}^{\dagger}(k)b_{\mu}(k) 
\\
& \quad + \sum_{\mu\neq \nu} V_{\mu,\nu}\int \frac{\dd{k}}{\sqrt{2\pi}}\int \frac{\dd{k'}}{\sqrt{2\pi}}b_{\mu}^{\dagger}(k)b_{\nu}(k')
\\
& \quad + \sum_{\mu} \int \dd{k}\qty(\frac{\kappa_{\mu}}{\sqrt{2\pi}}\sigma_{+} b_{\mu}(k) + \mathrm{h.c.}),\label{eq:H_s_wg_TLE}
\end{split}
\end{align}
which is similar to Eq.~\eqref{eq:Hprime_a}, except here we have accounted for an arbitrary number of waveguide modes. The total lifetime of the emitter is $ 1/\sum_{\mu} |\kappa_{\mu}|^2 $. The relevant input-output equations are (see App.~\ref{app:smatrix})
\begin{subequations}
	\begin{align}
		\dot{\sigma}_{-}(t) &= A\sigma_{-}(t) + \mathbf{B}\mathbf{b}_{\mathrm{in}}(t) + \hat{f}(t) \label{eq:inputOutput_TLE_a},
		\\
		\mathbf{b}_{\mathrm{out}}(t) &= \mathbf{C}\mathbf{b}_{\mathrm{in}}(t) + \mathbf{D}\sigma_{-}(t) \label{eq:inputOutput_TLE_b},
	\end{align}
\end{subequations}
with
\begin{subequations}
	\begin{align}
		A &= -i\omega_{\mathrm{e}} -\frac{1}{2}\boldsymbol{\kappa}^{\dagger}\mathbf{G}^{-1} \boldsymbol{\kappa},
		\\
		\mathbf{B} &= - i\boldsymbol{\kappa}^{\dagger}\mathbf{G}^{-1},
		\\
		\mathbf{C} &= \mathbf{G}^* \mathbf{G}^{-1},
		\\
		\mathbf{D} &= -i\mathbf{G}^{-1} \boldsymbol{\kappa},
		\\
		\hat{f}(t) &=  2i \boldsymbol{\kappa}^{\dagger}\mathbf{G}^{-1}\sigma_{+}(t)\sigma_{-}(t) \mathbf{b}_{\mathrm{in}}(t), \label{eq:TLE_inOutCoeff_f}
		\\
		\mathbf{G} &= \qty(\mathds{1}+\frac{i}{2}\mathbf{V}).
	\end{align}
\end{subequations}
Inserting the input-output relations, Eq.~\eqref{eq:inputOutput_TLE_a}-\eqref{eq:inputOutput_TLE_b}, in Eq.~\eqref{eq:S_pk_1ph}, we find
\begin{align}
	S_{p;k}^{\mu;\nu} = C_{\mu;\nu}\delta(p-k) + D_{\mu}\mathcal{G}^{\nu}_{p;k},\label{eq:S_TLE_index_01}
\end{align}
with 
\begin{align}
	\mathcal{G}^{\nu}_{p;k}  = \mel{\varnothing}{\tilde{\sigma}_{-}(p)}{k^+}_{\nu} = i\qty(p-iA)^{-1}B_{\nu} \, \delta(p-k). \label{eq:G_mu_TLE}
\end{align}
Eq.~\eqref{eq:S_TLE_index_01} is found by Fourier transforming Eq.~\eqref{eq:inputOutput_TLE_a} and using $  \langle \varnothing |\hat{f}=0 $. Eq.~\eqref{eq:TLE_inOutCoeff_f} is derived using $ [\sigma_{-},\sigma_{+}] = \mathds{1}-2\sigma_{+}\sigma_{-} $ but would have yielded the same result had we assumed $ [\sigma_{-},\sigma_{+}] \approx \mathds{1} $. This is sometimes called the low-excitation approximation~\cite{rephaeli2012fewPhoton_IEEE} but in fact gives the exact result regardless, for single-photon scattering. As we will soon see, the low-excitation approximation is also exact for the single-photon case if we replace the TLE with a JC system.

In matrix form, Eq.~\eqref{eq:S_TLE_index_01} has the simple expression
\begin{align}
	\mathbf{S}_{p;k} = \qty[\mathbf{C} + i \frac{\mathbf{D}\mathbf{B}}{p-iA}]\delta(p-k) \label{eq:S1_TLE}.
\end{align}
Although we have accounted for an arbitrary number of waveguide modes, we take the PTE to affect only the transmission between two modes, here named right ($ R $) and left ($ L $), i.e.
\begin{align}
	V_{\mu,\nu} = \begin{cases}
		V & \text{if}\, \qty(\mu,\nu)=\qty(R, L) \,\text{or}\, \qty(\mu,\nu)=\qty(L,R)
		\\
		0 & \text{otherwise}.
	\end{cases}
\end{align}
Physically, this corresponds to a waveguide geometry where scattering into dissipative channels is only possible via the system and not via the PTE. In the special case where the TLE couples symmetrically to two waveguide channels ($ R $ and $ L $), with $ \kappa_{R}=\kappa_{L}=\sqrt{\Gamma/2} $, and also couples to a dissipation channel ($ D $), with $ \kappa_{D}=\sqrt{\gamma} $, we find
\begin{align}
	\mathbf{S}_{p;k} = \mqty(t(p) & r(p) \\ r(p) & t(p)) \delta(p-k),
\end{align}
where the transmission and reflection coefficients are
\begin{subequations}
	\begin{align}
		t(p) &=  t_B - \frac{i\frac{\tilde{\Gamma}^2}{2\Gamma}}{\qty(p-\omega_{\mathrm{e}})+i\qty(\tilde{\Gamma}+\gamma)/2},
		\\
		r(p) &= r_B - \frac{i\frac{\tilde{\Gamma}^2}{2\Gamma}}{\qty(p-\omega_{\mathrm{e}})+i\qty(\tilde{\Gamma}+\gamma)/2},
	\end{align}
\end{subequations}
respectively. Here, $ \tilde{\Gamma}=\Gamma/\qty(1+iV/2) $, as in Sec.~\ref{sec:singleExcitationDynamics_TLE}, which reduces to the form found in~\cite{rephaeli2013dissipation_photResearch} in the absence of a PTE.

\begin{figure*}[htp]
	\centering
	\includegraphics[width=\textwidth]
	{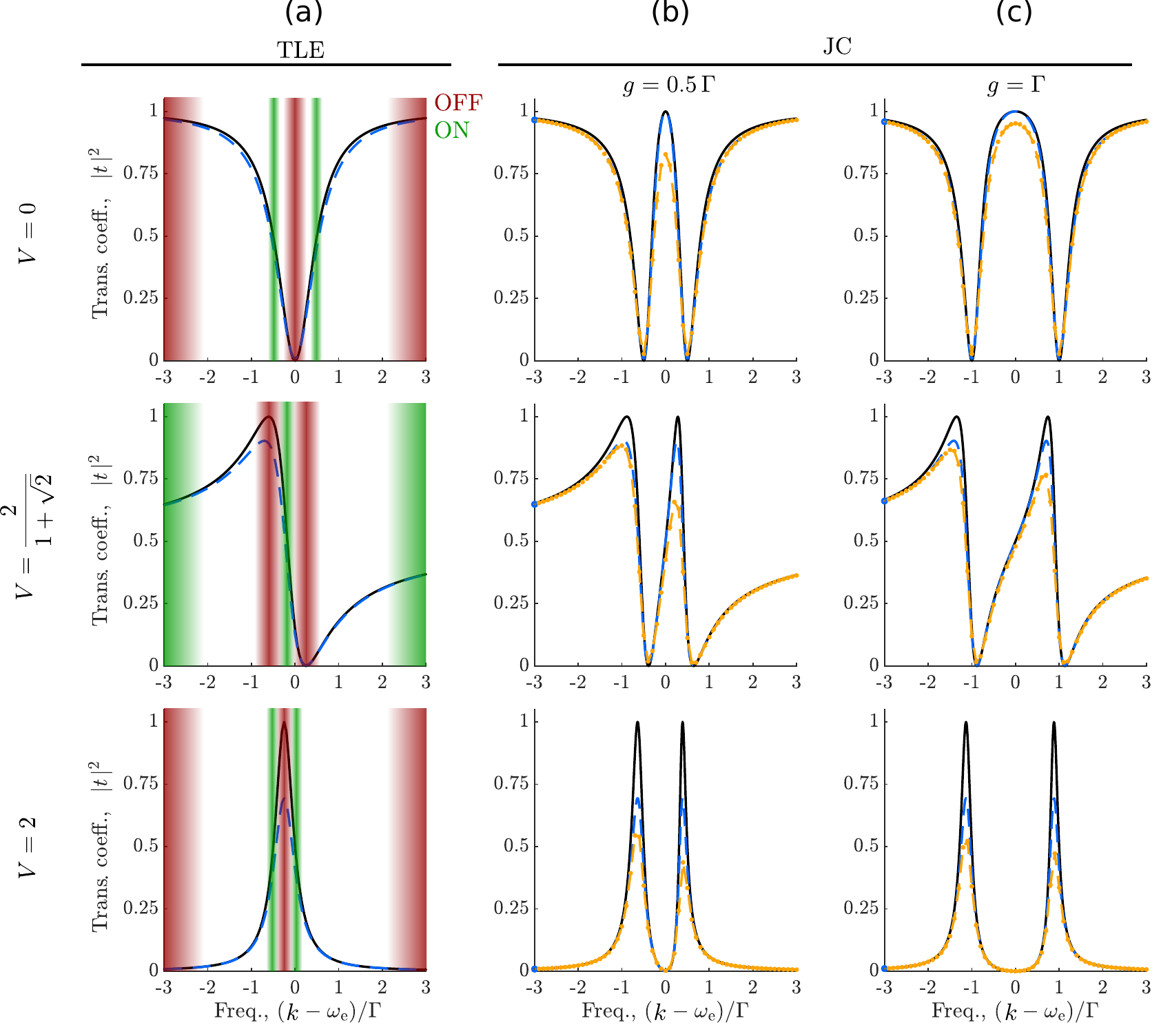}
	\caption{Absolute squared transmission coefficient, $ |t(k)|^2 $, for Fano-waveguide geometries with \textbf{(a)} a TLE and \textbf{(b)}-\textbf{(c)} a JC system with $ g=0.5\,\Gamma $ and $ g=1\,\Gamma $, respectively. The cavity energy is chosen such that $ \omega_{\mathrm{c}}=\omega_{\mathrm{e}} - \mathrm{Im}\{\tilde{\Gamma} \}/2 $. The PTE strengths are $ V=0 $ (row 1), $ V=2/\qty(1+\sqrt{2}) $ (row 2), and $ V=2  $ (row 3). In column \textbf{(a)}, the dissipation rates are $ \gamma = 0 $ (full black) and $ \gamma = 0.1 \,\Gamma $ (dashed blue). See Sec.~\ref{sec:switch} for an explanation of the red and green regions. In column (b) and (c), the dissipation rates are $ \qty(\gamma_{\mathrm{c}}, \gamma_{\mathrm{e}}) = \qty(0, 0)$ (full black), $ \qty(\gamma_{\mathrm{c}}, \gamma_{\mathrm{e}}) = \qty(0.1\,\Gamma, 0 )$ (dashed blue), and  $ \qty(\gamma_{\mathrm{c}}, \gamma_{\mathrm{e}}) = \qty(0.1\,\Gamma, 0.1\,\Gamma )$ (dashed dotted orange).
	}	\label{fig:transCoeff_lorentz_Fano_TLE_dissipation}
\end{figure*}

In Fig.~\ref{fig:transCoeff_lorentz_Fano_TLE_dissipation}(a) we plot the absolute squared transmission coefficient for the Fano-waveguide coupled to a TLE. We have chosen to illustrate three different PTE strengths both without ($ \gamma=0\, \Gamma $) and with ($ \gamma=0.1\, \Gamma $) dissipation. Without a PTE, row 1 of Fig.~\ref{fig:transCoeff_lorentz_Fano_TLE_dissipation}(a), the transmission spectrum is a simple Lorentzian lineshape. The dip in transmission is the result of destructive interference between the photon path of direct transmission and the transmission path via the emitter ~\cite{roulet2016two_PRA}.  A single incoming quasi-monochromatic photon resonantly tuned to a lossless emitter will thus reflect. Even when we account for dissipation, the dip is quite pronounced. For a balanced PTE, row 2 of Fig.~\ref{fig:transCoeff_lorentz_Fano_TLE_dissipation}(a), the transmission spectrum has a characteristic asymmetric Fano lineshape. Light incident from one end of the waveguide is transmitted due to partial transmission via the PTE and via the emitter. When these two paths add constructively, an increase in transmission is seen and vice versa; destructive interference results in a decrease in transmission. This gives a finite spectral distance between the maximum and minimum value in the spectrum. For a fully blocking PTE, row 3 of Fig.~\ref{fig:transCoeff_lorentz_Fano_TLE_dissipation}(a), the lineshape is again Lorentzian, but flipped upside-down relative to row 1. When we account for dissipation, the contribution from loss is more pronounced when high transmission coincides with the effective emitter resonance frequency. This is a consequence of light transmitted via the emitter being more susceptible to loss as only the emitter is assumed to couple to a dissipation channel. 

\subsection{Single-photon $ S $-matrix -- coupled cavity-emitter system}
We now determine the single-photon $ S $-matrix for the Fano-waveguide geometry coupled to a cavity-emitter system, as sketched in Fig.~\ref{fig:sketch_waveguide_PTE}(a), where $ \mathcal{S} $ is replaced by quantum system iii. of Fig.~\ref{fig:sketch_waveguide_PTE}(d). In the rotating wave approximation, the Hamiltonian is
	\begin{align}
	\begin{split}
	H &=\omega_{\mathrm{c}}a^{\dagger}a + \omega_{\mathrm{e}}\sigma_{+}\sigma_{-} + g\qty(\sigma_{-} a^{\dagger} + \sigma_{+} a)
	\\
	& \quad  + \sum_{\mu}\int \dd{k}\, k b_{\mu}^{\dagger}(k)b_{\mu}(k) 
	\\
	& \quad + \sum_{\mu\neq \nu} V_{\mu,\nu}\int \frac{\dd{k}}{\sqrt{2\pi}}\int \frac{\dd{k'}}{\sqrt{2\pi}}b_{\mu}^{\dagger}(k)b_{\nu}(k')
	\\
	& \quad + \sum_{\mu} \int \dd{k}\qty(\frac{\kappa_{\mathrm{c},\mu}a^{\dagger} + \kappa_{\mathrm{e},\mu}\sigma_{+}}{\sqrt{2\pi}} b_{\mu}(k) + \mathrm{h.c.}).\label{eq:H_s_cav_em}
	\end{split}
	\end{align}
Here, $ a $($ a^{\dagger} $) is the annihilation(creation) operator for the cavity and satisfy $ \comm{a}{a^{\dagger}}=1 $. $ g $  is the emitter-cavity coupling strength which we take to be real. Compared to Eq.~\eqref{eq:H_s_wg_TLE}, we have introduced an extra direction-dependent coupling coefficient for the cavity, $ \kappa_{\mathrm{c},\mu} $. We define a system operator vector, $ \hat{\mathbf{c}}=\mqty(a & \sigma_{-})^{T} $. The relevant input-output equations are (see App.~\ref{app:smatrix})
\begin{subequations}
	\begin{align}
		\dot{\mathbf{c}}(t) &= \mathbf{A}\mathbf{c}(t)+\mathbf{B}\mathbf{b}_{\mathrm{in}}(t) + \mqty(0 & \hat{f}(t))^T,
		\\
		\mathbf{b}_{\mathrm{out}}(t) &= \mathbf{C}\mathbf{b}_{\mathrm{out}}(t) + \mathbf{D}\mathbf{c}(t),
	\end{align}
\end{subequations}
with
\begin{subequations}
	\begin{align}
		\mathbf{A} &= \mqty(-i\omega_{\mathrm{c}} & -ig \\ -ig & -i\omega_{\mathrm{e}}) -\frac{1}{2}\boldsymbol{\kappa}^{\dagger} \mathbf{G}^{-1} \boldsymbol{\kappa},
		\\
		\mathbf{B}&=-i \boldsymbol{\kappa}^{\dagger}\mathbf{G}^{-1},
		\\
		\mathbf{C} &= \mathbf{G}^* \mathbf{G}^{-1},
		\\
		\mathbf{D} &= -i\mathbf{G}{-1} \boldsymbol{\kappa},
		\\
		\begin{split}
		\hat{f} &= \qty(2ig + \mqty(0&1)\boldsymbol{\kappa}^{\dagger}\mathbf{G}^{-1}\boldsymbol{\kappa}\mqty(1\\0) ) \sigma_{+}(t)\sigma_{-}(t) a(t) \\
		& \qquad +2i \mqty(0&1)\boldsymbol{\kappa}^{\dagger}\mathbf{G}^{-1}\sigma_{+}(t)\sigma_{-}(t) \mathbf{b}_{\mathrm{in}}(t) ,
		\end{split}
		\\
		\mathbf{G} &= \qty(\mathds{1}+\frac{i}{2}\mathbf{V}).
	\end{align}
\end{subequations}
Aside from an extra added dimension from having two independent system operators, these expressions look very similar to that of just the TLE, with $ \hat{f} $ again satisfying $  \bra{\varnothing}\hat{f}=0 $. The single-photon $ S $-matrix for this system is then found in a similar manner as for the case of the TLE. We find
\begin{align}
	\mathbf{S}^{(1)}_{p;k} = \qty[\mathbf{C} + i \mathbf{D}\qty(\mathds{1}p-i\mathbf{A})^{-1}\mathbf{B}]\delta(p-k) \label{eq:S1_JC}.
\end{align}

As before, we determine the transmission and reflection coefficients in a specific case. We assume that the cavity couples symmetrically to two waveguide channels ($ R $ and $ L $), with $ \kappa_{\mathrm{c},R}=\kappa_{\mathrm{c},L}=\sqrt{\Gamma/2} $, as well as to a dissipation channel ($ D_{\mathrm{c}} $), with $ \kappa_{\mathrm{c},D_{\mathrm{c}}}=\sqrt{\gamma_{\mathrm{c}}} $. Furthermore, we assume that the emitter couples only to the dissipation channel $ D_{\mathrm{e}} $ with strength $ \kappa_{\mathrm{e},D_{\mathrm{e}}}=\sqrt{\gamma_{\mathrm{e}}} $, i.e. $ \kappa_{\mathrm{e},R}=\kappa_{\mathrm{e},L}=0 $. The system now corresponds to a lossy JC system coupled to a Fano-waveguide. We find for the transmission and reflection coefficients:
\begin{subequations}
	\begin{align}
		t(p) &=  t_B - \frac{i \qty(p-\omega_{\mathrm{e}}+i\frac{\gamma_{\mathrm{e}}}{2})\frac{\tilde{\Gamma}^2}{2\Gamma}}{\qty[p-\omega_{\mathrm{c}}+i\frac{\tilde{\Gamma}+\gamma_{\mathrm{c}}}{2}]\qty[p-\omega_{\mathrm{e}}+i\frac{\gamma_{\mathrm{e}}}{2}]-g^2},
		\\
		r(p) &= r_B - \frac{i \qty(p-\omega_{\mathrm{e}}+i\frac{\gamma_{\mathrm{e}}}{2})\frac{\tilde{\Gamma}^2}{2\Gamma}}{\qty[p-\omega_{\mathrm{c}}+i\frac{\tilde{\Gamma}+\gamma_{\mathrm{c}}}{2}]\qty[p-\omega_{\mathrm{e}}+i\frac{\gamma_{\mathrm{e}}}{2}]-g^2}.
	\end{align}
\end{subequations}
In the limit $ g=0 $, we see that the scattering coefficients reduce to the case of the TLE with $ \omega_{\mathrm{c}} \rightarrow \omega_{\mathrm{e}} $ and $ \gamma_{\mathrm{c}} \rightarrow \gamma_{\mathrm{e}} $ as expected due to the low-excitation approximation. Furthermore, we note that the scattering coefficients are in agreement with~\cite{shen2009theory_1_PRA} for $ V=0 $.

In Fig.~\ref{fig:transCoeff_lorentz_Fano_TLE_dissipation}(b) and (c) we plot the absolute squared transmission coefficient for three different sets of dissipation rates using $ g=0.5 \, \Gamma $ and $ g=1\, \Gamma $, respectively. We choose the cavity frequency to be $ \omega_{\mathrm{c}}=\omega_{\mathrm{e}} - \mathrm{Im}\{\tilde{\Gamma} \}/2  $. As in the example of the TLE, row 1, 2, and 3 corresponds to having no PTE, a balanced PTE, and a fully blocking PTE in the waveguide, respectively. Without the PTE (row 1) the two dips in the transmission spectrum occur at the eigenfrequencies of the JC system, i.e. $ k =\omega_{\mathrm{e}}\pm g $. Qualitatively, column (b) and (c) are very similar. It is clear, however, that as $ g $ increases the lineshape will approach two copies of the TLE spectrum (Fig.~\ref{fig:transCoeff_lorentz_Fano_TLE_dissipation}(a) separated by $ 2g $. As expected, the effect of dissipation is again largest when a transmission peak coincides with one of the system eigenfrequencies. When the photon can escape via the emitter as well as via the cavity, the loss increases.

\subsection{Two-photon S-matrices}
The two-photon $ S $-matrix is calculated in a similar way as Eq.~\eqref{eq:S_pk_1ph}, except now with an additional photon before and after scattering. We denote the input directions $ \nu_1 $ and $ \nu_2 $, and the input frequencies $ k_1 $ and $ k_2 $. The output directions are $ \mu_1 $ and $ \mu_2 $, and the output frequencies $ p_1 $ and $ p_2 $. The general expression for the two-photon $ S $-matrix is:
\begin{align}
	S&_{p_1p_2;k_1k_2}^{\mu_1\mu_2;\nu_1\nu_2} = {}_{\mu_1\nu_2}\braket{p_1p_2^-}{k_1k_2^+}_{\nu_1\nu_2} \nonumber
	\\
	&=
	S_{p_1;k_1}^{\mu_1; \nu_1}S_{p_2;k_2}^{\mu_2; \nu_2}  +  S_{p_1;k_2}^{\mu_1; \nu_2} S_{p_2;k_1}^{\mu_2; \nu_1}   + iT_{p_1p_2;k_1k_2}^{\mu_1\mu_2;\nu_1\nu_2}. \label{eq:twoPhotonSmatrix}
\end{align}
The first two terms are the contributions from linear interactions. The third term is called the bound state term and accounts for the nonlinear interactions. For systems with a unique ground state and symmetrical waveguide coupling, the $ S $-matrix can be shown to have the following form \cite{xu2013analytic_PRL,xu2017generalized_PRA}
\begin{align}
	iT_{p_1p_2;k_1k_2}^{\mu_1\mu_2;\nu_1\nu_2} = \mathcal{M}_{p_1,p_2,k_1,k_2}\delta(p_1+p_2-k_1-k_2), \label{eq:boundStateTerms}
\end{align}
The delta function ensures energy conservation in the system. For a Fano-waveguide symmetrically coupled to a lossless TLE, we find (see App.~\ref{app:smatrix_TLE})
\begin{align}
	\mathcal{M}_{p_1,p_2,k_1,k_2} = \frac{1}{\pi}\frac{\tilde{\Gamma}}{\sqrt{2\Gamma}}\mathcal{G}(p_1)\mathcal{G}(p_2)\qty[\mathcal{G}(k_1)+\mathcal{G}(k_2)], \label{eq:S_TLE_nonlinearTerm}
\end{align}
with
\begin{align}
	\mathcal{G}(k) = \frac{\frac{\tilde{\Gamma}}{\sqrt{2\Gamma}}}{k-\omega_{\mathrm{e}}+i\frac{\tilde{\Gamma}}{2} }.
\end{align}
Replacing the TLE for a JC system, we find (see App.~\ref{app:smatrix_JC})
\begin{align}
	\mathcal{M}_{p_1,p_2,k_1,k_2} &=\frac{-2ig}{\sqrt{2\pi}} \mathcal{G}_{\mathrm{e}}(p_1) \mathcal{G}_{\mathrm{e}}(p_2) \mathcal{G}_{\mathrm{e-c}}(p_1+p_2,k_1,k_2) ,\label{eq:S_JC_nonlinearTerm}
\end{align}
with
\begin{widetext}
	\begin{align}
		\mathcal{G}_{\mathrm{e-c}}(p_1+p_2,k_1,k_2)    &= - \frac{\frac{\tilde{\Gamma}/2}{\sqrt{\Gamma/2}}}{\sqrt{2\pi}} \frac{ \qty[p_1+p_2-2\omega_{\mathrm{c}} +i \tilde{\Gamma}]\qty(\mathcal{G}_\mathrm{e}(k_1)+\mathcal{G}_\mathrm{e}(k_2)) + 2g \qty(\mathcal{G}_\mathrm{c}(k_1)+\mathcal{G}_\mathrm{c}(k_2))  }{ \qty[p_1+p_2 - \qty(\omega_{\mathrm{c}} + \omega_{\mathrm{e}} )+i\frac{\tilde{\Gamma}}{2}] \qty[p_1+p_2-2\omega_{\mathrm{c}} +i \tilde{\Gamma}] -2g^2 } ,
	\end{align}
	and
	\begin{align}
		\mqty(\mathcal{G}_{\mathrm{c}}(k) \\  \mathcal{G}_{\mathrm{e}}(k)) = \frac{1}{\qty[k- \omega_{\mathrm{c}} +i \frac{\tilde{\Gamma} }{2}]\qty[k -\omega_{\mathrm{e}} ] - g^2} \mqty( \frac{\tilde{\Gamma}/2}{\sqrt{\Gamma/2}} \qty(k-\omega_{\mathrm{e}})  \\  \frac{\tilde{\Gamma}/2}{\sqrt{\Gamma/2}} g ).
	\end{align}
\end{widetext}

Going beyond two-photon scattering can be achieved by an expansion of the expression in Eq.~\eqref{eq:twoPhotonSmatrix}, see eg.~\cite{xu2015inputOutput_PRA}. The caveat, however, is the fast increase in the number of terms one has to consider. We therefore limit ourselves to the analysis of single and two-photon scattering properties in the next section.
%%%%%%%%%%%%%%%%%%%%%%%%%% HOM SWITCHING %%%%%%%%%%%%%%%%%%%%%%%%%
%%%%%%%%%%%%%%%%%%%%%%%%%%%%%%%%%%%%%%%%%%%%%%%%%%%%%%%%%%%%%%%%%%
\section{Hong-Ou-Mandel switch}\label{sec:switch}
It has recently been experimentally demonstrated that the Fano effect can be used to make a tunable quantum optical filter in a waveguide~\cite{hallett2019tunable_PRL}. Recent theoretical results have shown that the asymmetry of the Fano resonance can be used to suppress electron-phonon interactions in quantum dots~\cite{denning2019quantum_PRB}. In this section, we investigate how the Fano effect may be used to implement a tunable Hong-Ou-Mandel (HOM) switch. As was suggested by Roulet et al.~\cite{roulet2016two_PRA}, with circulators in optical waveguides and the ability to efficiently turn the HOM interference on and off, a configurable integrated on-chip circuit can be realized. As an example, they analyzed the scattering behavior of two initially counter-propagating photons in a waveguide coupled to a TLE. In this section we extend the analysis to include a PTE element in the waveguide, and investigate how the Fano interference effect may be exploited. We will assume that the two incident photons are identical, with Gaussian envelopes, i.e.
\begin{align}
	\ket{LR} = \int \dd{k_L} \xi(k_L)b_{L}^{\dagger}(k_L)\int \dd{k_R} \xi(k_R)b_{R}^{\dagger}(k_R)\ket{\varnothing},
\end{align}
where $ \xi(k) $ is the same spectral envelope as in Eq.~\eqref{eq:Gaussian}. We define an ideal HOM switch as having the following properties
\begin{align}
	\ket{LR} \longrightarrow \begin{cases}
		\ket{LR} & \text{switch-OFF},
		\\
		\frac{1}{\sqrt{2}}\qty[\ket{LL} + \ket{RR}] & \text{switch-ON},
	\end{cases}
\end{align}
where
\begin{align}
	\ket{LL(RR)} = \frac{1}{\sqrt{2}}\qty[\int \dd{k} \xi(k)b_{L(R)}^{\dagger}(k)]^2\ket{\varnothing}.
	\label{eq:HOM_ket_LL_RR}
\end{align}
In the switch-OFF configuration, the output state, just as the input, represents two photons propagating in different channels, whereas in the ON configuration, the output is a superposition of two photons propagating in the same channel. The latter is equivalent to a beam splitter transformation, except here the output arms are mapped onto the input arms. 

\subsection{Quasi monochromatic input photons}
We first consider quasi monochromatic photons, i.e. the limit where $ \sigma\rightarrow 0 $ (See Eq.~\eqref{eq:Gaussian}). Because we have assumed the photons to be quasi monochromatic, correlations arising from the nonlinear term $ \mathcal{M}_{p_1,p_2,k_1,k_2} $ (Eq.~\eqref{eq:S_TLE_nonlinearTerm}) in the scattering matrix effectively vanishes \cite{nysteen2015scattering_NJoP}. In other words, the scattering event is linear, and the resulting output state is found by considering the two photons independently. This means, the switch is in the OFF configuration, i.e. the output state equals the input state $ |LR\rangle $, when the transmission coefficient is either 0 or 1, as this corresponds to both photons being perfectly reflected or transmitted, respectively. Conversely, the ON-configuration is reached when the absolute squared transmission coefficient is $ 1/2 $. In this case, each photon scatters into an equal superposition of being transmitted and reflected, but the terms corresponding to a counter-propagating output cancel. From the spectra in Fig.~\ref{fig:transCoeff_lorentz_Fano_TLE_dissipation}(a) we can therefore easily identify the ON and OFF configurations indicated by the green and red regions, respectively. For $ V=0 $ (row 1) the OFF configuration is reached when the two photons are either far detuned from the emitter, or when the two photons both are resonant with the emitter. In the former case, each photon will transmit without interaction and in the latter case, the two photons will both reflect as if there were a mirror. For the Lorentzian spectrum in row 1, the ON configuration is reached when the photon-emitter detuning is $ \omega = \pm \Gamma/2  $. 

An unaddressed caveat for the protocol by Roulet et al. \cite{roulet2016two_PRA} is that in order to realize the ON-configuration in the absence of a PTE ($ V=0 $), the emitter must be tuned to the energy where the transmission lineshape is the steepest, making the protocol highly susceptible to errors. In case of a balanced PTE, the ON-configuration is reached when the emitter-photon detuning is large compared to the waveguide coupling rate, $ \Gamma $, or when $ k_c-\omega_{\mathrm{e}}=\tilde{\omega}_{\mathrm{e}} $ (see row 2 of Fig.~\ref{fig:transCoeff_lorentz_Fano_TLE_dissipation}(a)).  The OFF-configuration is reached for $ k_c-\omega_{\mathrm{e}} = \tilde{\omega}_{\mathrm{e}} \pm  (2+\sqrt{2})\Gamma/8$. Both the ON and OFF-configurations can thus be realized at frequencies where the derivative of the transmission coefficient is (arbitrarily close to) 0, thus suggesting that introducing a balanced PTE in the waveguide results in a HOM switch that is more robust with respect to small perturbations in both the emitter and photon energy. The price of achieving this, however, is that it requires an emitter detuning of several linewidths. For a PTE strength of $ V=2 $, the transmission spectrum is again Lorentzian, with ON-configurations at points where the slope is steep (row 3 of Fig.~\ref{fig:transCoeff_lorentz_Fano_TLE_dissipation}(a)).

The analysis of the TLE can be carried over to the JC system: The OFF-configuration is reached for a transmission coefficient of zero or unity, whereas the ON-configuration is reached when the absolute squared transmission coefficient equals $ 1/2 $. For the balanced PTE, this again means that both the ON-and OFF-configurations can be reached at points where the slope is zero (see  Fig.~\ref{fig:transCoeff_lorentz_Fano_TLE_dissipation}(b) and (c)).

\subsection{Finite-width envelopes}
\begin{figure*}[htp]
	\centering
	\includegraphics[width=\textwidth]
	{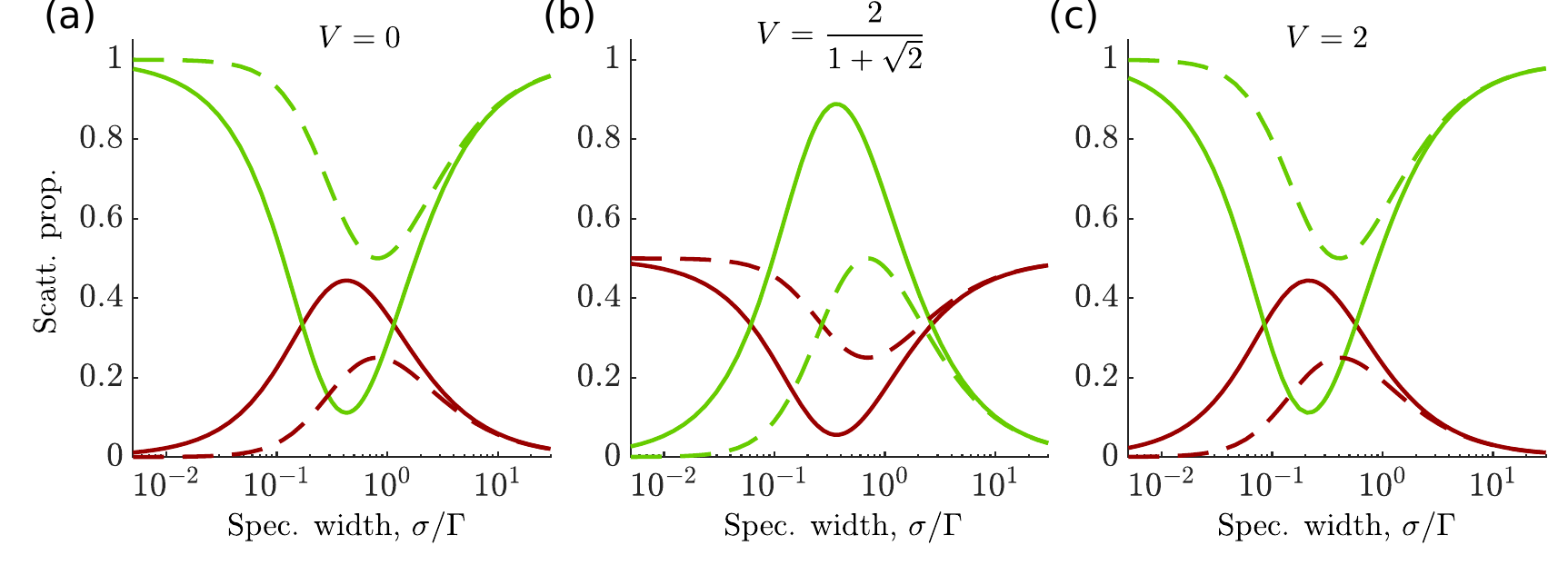}
	\caption{Scattering probabilities for two counter-propagating photons on a TLE versus spectral width. The probability of counter-propagating output photons are shown in green and the probability for co-propagating output photons are shown in red. The full line shows the result for scattering off a TLE, whereas the dashed line shows the result for scattering off a cavity, i.e. without the bound state contribution in the $ S $-matrix. The envelopes are Gaussian with $ k_c = \tilde{\omega}_{\mathrm{e}} $. The PTE strengths are {\bf (a)} $ V=0 $, {\bf (b)} $ V=2/(1+\sqrt{2}) $ and {\bf (c)} $ V=2 $. }	\label{fig:switch}
\end{figure*}

As a final investigation, we consider the scattering of counter-propagating photons with finite spectral width in Fano-resonance waveguides. As has been shown for the case of a TLE in a waveguide with $ V=0 $, strong scattering-induced photon correlations are possible for envelope widths comparable to the waveguide coupling rate~\cite{nysteen2015strong_PRA}. In this connection, it was also suggested how the TLE can be used as a beam splitter for photons with envelopes of finite width. For Gaussian envelopes centered around $ \omega_{\mathrm{e}} $, the scattering properties depend highly on the envelope width $ \sigma $ (see Fig.~\ref{fig:switch}(a)). For spectrally narrow or broad envelopes, two counter propagating photons will continue to be counter-propagating post scattering, i.e. the OFF-configuration is achieved. The reason for the former was explained in the previous subsection. The reason for the latter can most easily be explained by considering the wide distribution in frequencies; the photon is practically everywhere far-detuned from the emitter and thus does not interact with it. For $ \sigma = 0.43\, \Gamma $, the probability of a HOM-like interference is maximized. For this width the trade-off between having too narrow and too broad spectral envelopes is minimized. The ON-configuration is reached with an error probability (the probability of scattering into a counter-propagating output state, $ |LR\rangle $)), of $ 0.11 $.

In Fig.~\ref{fig:switch}(b), we plot the scattering probabilities versus envelope width in case of a balanced PTE, again for envelopes centered around $ \tilde{\omega_{\mathrm{e}}} $. In this case, the ON-configuration is reached for spectrally narrow or broad envelopes because only the PTE contributes to the scattering. The OFF configuration is maximized for $ \sigma = 0.36\, \Gamma $, also with an error probability (the probability of detecting two co-propagating photons) of $ 0.11 $. As seen from Fig.~\ref{fig:switch}(c), a PTE strength of $ V=2 $ gives qualitatively the same results as having no PTE. We note that a center frequency of $ \tilde{\omega}_{\mathrm{e}} $ was chosen as this maximizes the switching contrast (not shown here).

The dashed lines in Fig.~\ref{fig:switch} correspond to scattering off a cavity (system ii. in Fig.~\ref{fig:sketch_waveguide_PTE}(d)) in place of the TLE. This corresponds to setting the bound state term in Eq.~\eqref{eq:boundStateTerms} equal to zero, as can easily be derived using $ [a,a^{\dagger}] = \mathds{1}  $ in place of $ [\sigma_{-},\sigma_{+}] = \mathds{1}-2\sigma_{+}\sigma_{-} $ for the two-photon $ S $-matrix derivation (see Sec.~\ref{sec:sMatrix_TLE}). Because there are no nonlinear scattering effects for this system, the minimum error probability is $ 0.5 $ in all three cases.

\begin{figure*}[htp]
	\centering
	\includegraphics[width=\textwidth]
	{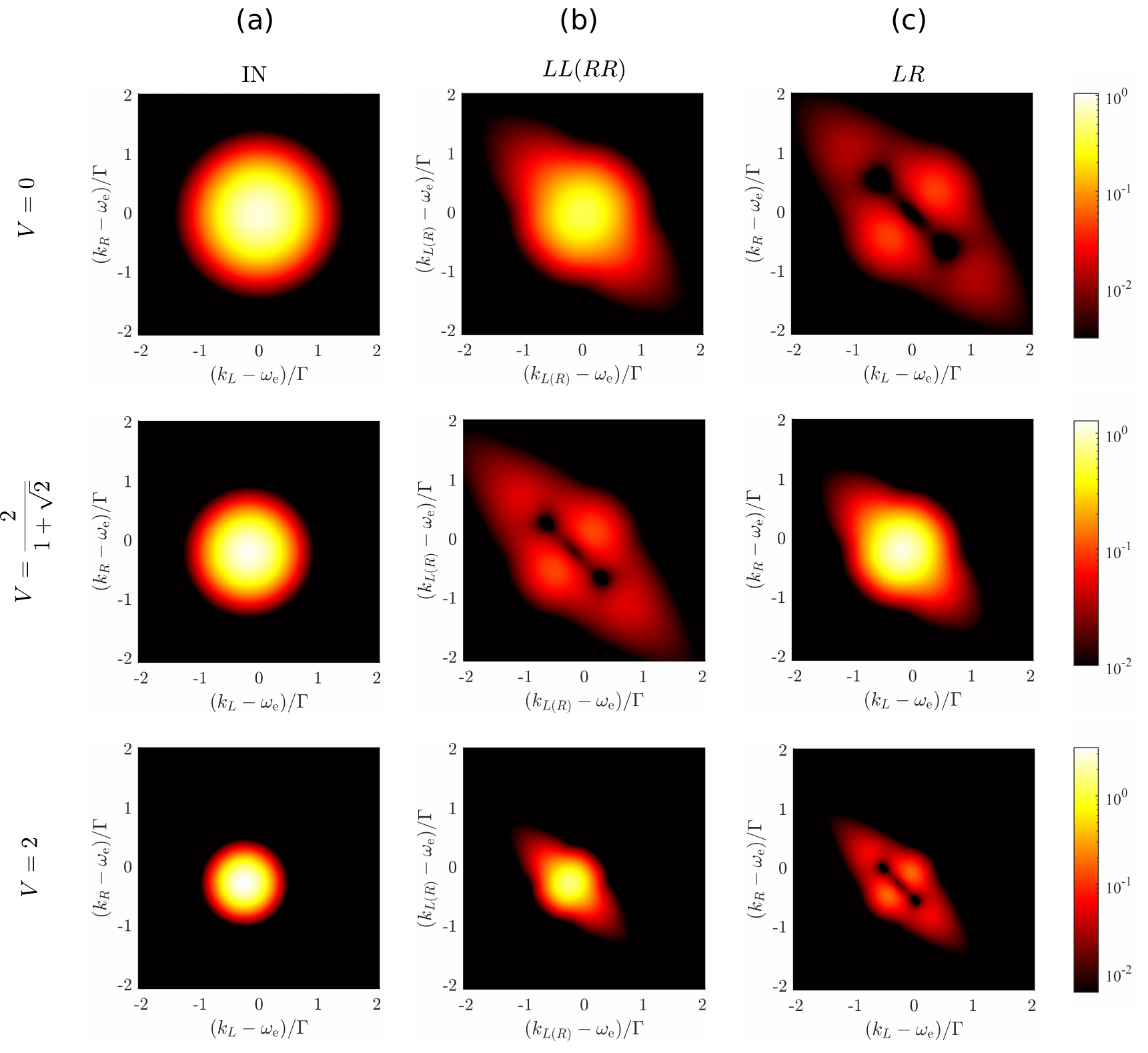}
	\caption{Two-photon envelope densities (color scale) vs. frequencies. The input state (column {\bf{(a)}}) is taken to be a counter-propagating two-photon state traveling towards a TLE in a waveguide with a PTE. The input photons have identical Gaussian envelopes with the center frequency $ k_c = \tilde{\omega}_{\mathrm{e}} $. In row 1 the PTE strength is $ V=0 $ and the spectral width is $ \sigma=0.43\Gamma $. In row 2 the PTE strength is $ V=2/(1+\sqrt{2})$ and the spectral width is $ \sigma=0.36\Gamma $. In row 3 the PTE strength is $ V=2$ and the spectral width is $ \sigma=0.21\Gamma $. Column {\bf{(b)}} shows the co-propagating output part ($ LL $ or $ RR $) and column {\bf{(c)}} 3 shows the counter-propagating output spectrum ($ LR $).}	\label{fig:twoPh_spec}
\end{figure*}

Finally, we plot in Fig.~\ref{fig:twoPh_spec} the two-photon input and output spectra for the envelope widths that minimize the error probabilities, defined above. Ideally, the input spectrum (fig.~\ref{fig:twoPh_spec}(a)) would be identical to the desired output spectrum (row 1 and row 3 of Fig.~\ref{fig:twoPh_spec}(b) for $ V=0 $ and $ V=2 $, respectively, and row 2 of Fig.~\ref{fig:twoPh_spec}(c) for $ V=2/(1+\sqrt{2}) $). We see, however, that the output spectra accumulates correlations in the spectra due to wave mixing arising from the nonlinear term in Eq.~\eqref{eq:S_TLE_nonlinearTerm}, resulting in elongated envelope densities. We see that the Fano resonance effect does not alter the induced correlations.
%%%%%%%%%%%%%%%%%%%%%%%%%% CONCLUSION  %%%%%%%%%%%%%%%%%%%%%%%%%%%
%%%%%%%%%%%%%%%%%%%%%%%%%%%%%%%%%%%%%%%%%%%%%%%%%%%%%%%%%%%%%%%%%%
\section{\label{sec:conclusion}Summary}
In this work we have analyzed the effect of Fano interference on few-photon scattering in waveguide geometries containing a partially transmitting element. In Sec.~\ref{sec:intro} we briefly motivated the study of such waveguide geometries, and in Sec.~\ref{sec:model} we described the model Hamiltonian for a waveguide with a partially transmitting element as introduced in~\cite{xu2016fano_PRA}. The work presented in this paper contributes to the study of Fano waveguide geometries in a quantum optical context by analyzing dynamical single-photon scattering, scattering of photons with finite-width envelopes, and providing analytical expressions for single and two-photon $ S $-matrices for more complicated waveguide geometries than studied previously. Below we give a brief summary of our results presented in Sec.~\ref{sec:singleExcitationDynamics_TLE} to Sec.~\ref{sec:switch}.

In Sec.~\ref{sec:singleExcitationDynamics_TLE} we found analytical expressions for the single-excitation dynamics for a two-level emitter in a waveguide with a partially transmitting element in the mirror-symmetry line. We showed that the emitter lifetime increases with the strength of the partially transmitting element, and that the emitted photons are redshifted. Furthermore, we determined the conditions for which a single-photon with Gaussian envelope maximizes the emitter excitation probability. In the limit of a vanishing partially transmitting element, we reproduce the results of e.g.~\cite{chen2011coherent_NJoP} and~\cite{liao2015singlePhoton_PRA}.

In Sec.~\ref{sec:S-matrix} we established a way of calculating the single-photon $ S $-matrix for a general Fano-waveguide geometry including the effect of dissipation. We also determined the two-photon S-matrix for a two-level emitter and a Jaynes-Cummings system. In case of a TLE, we showed that the results agree with that of Xu and Fan~\cite{xu2016fano_PRA}. In case of a Jaynes-Cummings system, we obtain the same $ S $-matrix coefficients as~\cite{shen2009theory_1_PRA} in the limit of a vanishing partially transmitting element.

In Sec.~\ref{sec:switch} we discussed the switching properties of a Fano-waveguide geometry and compared the results to that of a waveguide without a partially transmitting element, as studied by Roulet et al. in~\cite{roulet2016two_PRA}. We showed that for quasi-monochromatic photons, the Fano interference effect can be exploited to generate tunable Hong-Ou-Mandel interference between counter-propagating photons in a way that is robust against small changes in the emitter energy. For finite width-envelopes, we investigated the Fano interference effect on the induced spectral correlations and found no advantage compared to the case without a Fano interference.

\appendix
\begin{widetext}
%%%%%%%%%%%%%%%%%%%%%%%%%% APP. A  %%%%%%%%%%%%%%%%%%%%%%%%%%%
%%%%%%%%%%%%%%%%%%%%%%%%%%%%%%%%%%%%%%%%%%%%%%%%%%%%%%%%%%%%%%%%%%
\section{Single-excitation dynamics in a Fano-waveguide geometry coupled to a TLE} \label{app:1excDyn_TLE}
The Hamiltonian for a TLE coupled to a linearly dispersive waveguide with a PTE is given by (see Eq.~\eqref{eq:Hprime_a})
\begin{align}
	\begin{split}
		H &= \omega_{\mathrm{e}}\sigma_{+}\sigma_{-} + \int \dd{k}\, k \qty(b_R^{\dagger}(k)b_R(k) +b_L^{\dagger}(k)b_L(k) ) + \int \dd{k}\, \frac{\sqrt{\Gamma/2}}{\sqrt{2\pi}}\qty(\sigma_{-}\qty[b_R^{\dagger}(k)+b_L^{\dagger}(k)] + \mathrm{h.c.})
		\\
		& \qquad   + V \int \frac{\dd{k}}{\sqrt{2\pi}}\frac{\dd{k'}}{\sqrt{2\pi}} \, \qty[ b_R^{\dagger}(k)b_L(k') + \mathrm{h.c.} ].
	\end{split}
\end{align}
The state-vector expansion in the single-excitation manifold is
\begin{align}
	\ket{\psi(t)} = \chi(t) \sigma_{+}\ket{\varnothing} + \int \dd{k}\, \xi_R(k,t) b_R^{\dagger}(k)\ket{\varnothing} + \int \dd{k}\, \xi_L(k,t) b_L^{\dagger}(k)\ket{\varnothing}. \label{eq:app_psi}
\end{align}
Inserting in the Schr{\"o}dinger equation, we find the following equations of motion:
\begin{subequations}
	\begin{align}
		i\partial_t \chi(t) &= \omega_e \chi(t)  +  \frac{\sqrt{\Gamma/2}}{\sqrt{2\pi}}\int \dd{k}\,  \qty[\xi_R(k,t) + \xi_L(k,t)],
		\\
		i\partial_t \xi_R(k,t) &= k\xi_R(k,t)   +  \frac{\sqrt{\Gamma/2}}{\sqrt{2\pi}}\chi(t) + \frac{V}{2\pi}\int \dd{k'}\, \xi_L(k',t), \label{eq:app:xi_R_00}
		\\
		i\partial_t \xi_L(k,t) &= k\xi_L(k,t)   +  \frac{\sqrt{\Gamma/2}}{\sqrt{2\pi}}\chi(t) + \frac{V}{2\pi}\int \dd{k'}\, \xi_R(k',t). \label{eq:app:xi_L_00}
	\end{align}
\end{subequations}
We define $ \xi(k,t) = \xi_R(k,t) + \xi_L(k,t) $ and make the variable change $ \chi(t) = \chi'(t) e^{-i\omega_{\mathrm{e}} t} $ and $ \xi(k,t) = \xi'(k,t)e^{-ik t} $:
\begin{subequations}
	\begin{align}
		i\partial_t \chi'(t) &=   \frac{\sqrt{\Gamma/2}}{\sqrt{2\pi}}e^{i\omega_{\mathrm{e}} t}\int \dd{k}\,  \xi'(k,t) e^{-ik t} \label{eq:app:chi_01},
		\\
		i\partial_t \xi'(k,t) &=    \frac{\sqrt{\Gamma}}{\sqrt{\pi}}\chi'(t) e^{-i\qty(\omega_{\mathrm{e}}-k)t}  + \frac{V}{2\pi}\int \dd{k'}\, \xi'(k',t)e^{i\qty(k -k')t} \label{eq:app:xi_01}.
	\end{align}
\end{subequations}
We formally integrate Eq.~\eqref{eq:app:xi_01} from $ t_i $ to $ t $ and then multiply by $-i e^{-ik t} $:
\begin{align}
	\begin{split}
		\xi'(k,t)e^{-ik t} &= \xi'(k,t_i)e^{-ik t}   -i \frac{\sqrt{\Gamma}}{\sqrt{\pi}}\int_{t_i}^{t} \dd{t'}\,\chi'(t')e^{-ik \qty(t-t') }e^{-i\omega_{\mathrm{e}} t'}
		-i\frac{V}{2\pi}\int_{t_i}^{t} \dd{t'}\,  \int \dd{k'}\, \xi'(k',t') e^{-ik' t'}e^{ik \qty(t'-t)}.
	\end{split}
\end{align}
We integrate over all $ k $ and find
\begin{gather}
	\int \dd{k}\, \xi'(k,t)e^{-ik t} =\qty(1+\frac{iV}{2})^{-1} \int \dd{k}\,\xi'(k,t_i)e^{-ik t}   -i\qty(1+\frac{iV}{2})^{-1}\sqrt{\pi \Gamma} \chi'(t) e^{-i\omega_{\mathrm{e}} t} .
\end{gather}
We insert above expression in Eq.~\eqref{eq:app:chi_01}
\begin{align}
	\partial_t \chi'(t) &=-i \frac{\sqrt{\Gamma/2}}{\sqrt{2\pi}}\qty(1+\frac{iV}{2})^{-1} \int \dd{k}\,\xi'(k,t_i)e^{-i\qty(k -\omega_{\mathrm{e}}) t}   -\frac{\Gamma}{2}\qty(1+\frac{iV}{2})^{-1}  \chi'(t)  .
\end{align}
After formal integration we find
\begin{align}
	\chi(t) &= \chi(t_i)e^{-\qty[\frac{\tilde{\Gamma}}{2}+i\omega_{\mathrm{e}}]\qty(t-t_i)} - i\frac{\tilde{\Gamma}}{2\sqrt{\pi \Gamma}}
	\int \dd{k'}\, \xi(t_i,k')e^{-ik'\qty(t-t_i)}\frac{1-e^{-\qty[\frac{\tilde{\Gamma}}{2}+i\qty(\omega_{\mathrm{e}}-k')]\qty(t-t_i)}}{\frac{\tilde{\Gamma}}{2}+i\qty(\omega_{\mathrm{e}}-k')}\label{eq:app_chi},
\end{align}
where we have defined $ \tilde{\Gamma} = \Gamma/\qty(1+\frac{iV_{RL}}{2}) $. 
Next, we solve for $ \xi(k,t) $. Although this is not technically a photon envelope coefficient it is a helpful variable.
From combining Eq.~\eqref{eq:app:chi_01} and Eq.~\eqref{eq:app:xi_01} we find
\begin{align}
	\partial_t \xi'(k,t)  &= -ie^{ik t}\qty[\frac{\sqrt{\Gamma}}{\sqrt{\pi}}\qty(1 -\frac{V}{\Gamma}\omega_{\mathrm{e}})\chi(t) +i  \frac{V}{\sqrt{\pi\Gamma}}\partial_t \chi(t) ].
\end{align}
We solve by formal integration
\begin{align}
	\begin{split}
		\xi(t,k)&=\xi(t_i,k)e^{-ik \qty(t-t_i)}  -i \frac{\sqrt{\Gamma}}{\sqrt{\pi}}\qty(1 -\frac{V}{\Gamma}\omega_{\mathrm{e}}) \int_{t_i}^{t} \dd{t'} \, e^{ik \qty(t'-t)} \chi(t')  +   \frac{V}{\sqrt{\pi\Gamma}} \int_{t_i}^{t} \dd{t'} \,  e^{ik \qty(t'-t)} \partial_{t'}\chi(t'). \label{eq:app_xi}
	\end{split}
\end{align}
To solve for $ \xi_R(k,t) $ we use that
\begin{align}
	\xi_L'(t,k) &= \xi_L'(t_i,k)  -i \frac{\sqrt{\Gamma/2}}{\sqrt{2\pi}} \int_{t_i}^{t} \dd{t'}\, \chi'(t')e^{-i\qty(\omega_{\mathrm{e}}-k)t'}  -i \, \frac{V}{2\pi}  \int_{t_i}^{t} \dd{t'} \int \dd{k'}\, \xi_R'(t',k') e^{-i\qty(k'-k) t'},
\end{align}
and insert in Eq.~\eqref{eq:app:xi_R_00}. With this we find
\begin{align}
	\begin{split}
		\xi_R(t,k) &= \xi_R(t_i,k)e^{-ik\qty(t-t_i)}  -i\sqrt{\frac{\Gamma/2}{2\pi}}\frac{1-iV}{1-\frac{iV}{2}}\int_{t_i}^{t}\dd{t'}\, e^{ik(t'-t)}\chi(t')
		\\
		& \quad  -\frac{i}{2\pi}\frac{V}{1-i\frac{V}{2}} \int_{t_i}^{t}\dd{t'}\, e^{ik(t'-t)}\int \dd{k'}\, \xi_L(t_i,k')e^{-ik'\qty(t'-t_i)}  -\frac{1}{\pi}\frac{\qty(V/2)^2}{1-i\frac{V}{2}} \int_{t_i}^{t}\dd{t'}\, e^{ik(t'-t)}\int \dd{k'}\, \xi(t',k'). \label{eq:app_xi_R}
	\end{split}
\end{align}

To summarize, the dynamical evolution of the excitation probability is calculated from Eq.~\eqref{eq:app_chi}. The dynamical evolution of the right-propagating envelope is calculated using Eq.~\eqref{eq:app_xi_R}, where $ \xi $ is given in Eq.~\eqref{eq:app_xi}. To calculate the left-propagating envelope we simply use $ \xi_L(t,k) = \xi(t,k)-\xi_R(t,k) $. From this, all the state vector coefficients in Eq.~\eqref{eq:app_psi} are known provided we know the initial conditions $ \chi(t_i) $, $ \xi_R(t_i,k) $, and $ \xi_L(t_i,k) $ for some initial time $ t_i $.
%%%%%%%%%%%%%%%%%%%%%%%%%% APP. B  %%%%%%%%%%%%%%%%%%%%%%%%%%%
%%%%%%%%%%%%%%%%%%%%%%%%%%%%%%%%%%%%%%%%%%%%%%%%%%%%%%%%%%%%%%%%%%
\section{Single and two-photon $ S $-matrix calculations} \label{app:smatrix}
We start from the Hamiltonian in Eq.~\eqref{eq:H_s_cav_em} for a cavity-emitter system. We repeat the relevant terms here for easy reference:
\begin{subequations}
	\begin{align}
		\begin{split}
			H &= \sum_{\mu}\int \dd{k}\, k b_{\mu}^{\dagger}(k)b_{\mu}(k)  + \sum_{\mu\neq \nu} V_{\mu,\nu}\int \frac{\dd{k}}{\sqrt{2\pi}}\int \frac{\dd{k'}}{\sqrt{2\pi}}b_{\mu}^{\dagger}(k)b_{\nu}(k')
			+ H_{\mathrm{s}} + H_{\mathrm{s-wg}} \label{eq:app_H_general},
		\end{split}
		\\
		H_{\mathrm{s}} &= k_{\mathrm{c}}a^{\dagger}a + \omega_{\mathrm{e}}\sigma_{+}\sigma_{-} + g\sigma_{-} a^{\dagger} + g^* \sigma_{+} a, \label{eq:app_H_s_cav_em}
		\\
		H_{\mathrm{s-wg}} &= \sum_{\mu} \int \dd{k}\frac{\kappa_{\mathrm{c},\mu}a^{\dagger} + \kappa_{\mathrm{e},\mu}\sigma_{+}}{\sqrt{2\pi}} b_{\mu}(k) + \mathrm{h.c.}. \label{eq:app_H_sWG_cav_em}
	\end{align}
\end{subequations}
In the limit $ g=0 $, and $ \kappa_{\mathrm{c},\mu}=0 $ for all $ \mu $, this reduces to the Hamiltonian in Eq.~\eqref{eq:H_s_wg_TLE}.

\subsection{Input-output relations}
From the Heisenberg equation of motion we find for the waveguide operators
\begin{align}
	\partial_t b_{\mu}(k,t) &= -i \comm{b_{\mu}(k)}{H}  = -ik b_{\mu}(k,t) -i \frac{\kappa_{\mu,\mathrm{c}}}{\sqrt{2\pi}}a(t)-i \frac{\kappa_{\mu,\mathrm{e}}}{\sqrt{2\pi}}\sigma_{-}(t)-i\frac{\sum_{\nu \setminus \mu}V_{\mu \nu}\Phi_{\nu}(t)}{\sqrt{2\pi}}, \label{eq:app_dt_b_mu}
\end{align}
where
\begin{align}
	\Phi_{\nu}(t) = \frac{1}{\sqrt{2\pi}} \int \dd{k} \, b_{\nu}(k,t).
\end{align}
We rewrite Eq.~\eqref{eq:app_dt_b_mu}
\begin{align}
	\partial_t\qty{ e^{ik t} b_{\mu}(k,t) } &= -i \frac{\kappa_{\mu,\mathrm{c}}}{\sqrt{2\pi}}a(t) e^{ik t}-i \frac{\kappa_{\mu,\mathrm{e}}}{\sqrt{2\pi}}\sigma_{-}(t)  e^{ik t}-i\frac{\sum_{\nu \setminus \mu}V_{\mu \nu}\Phi_{\nu}(t)}{\sqrt{2\pi}}  e^{ik t}. \label{eq:app_dt_b_mu_rewrite}
\end{align}
We formally integrate from time $ t_{\mathrm{in}} $ to $ t $ and multiply through by $ e^{-ikt} $
\begin{align}
	\begin{split}
		b_{\mu}(k,t) -  e^{ik \qty(t_{\mathrm{in}} - t)} b_{\mu}(k,t_{\mathrm{in}})  &= -i \int_{t_{\mathrm{in}}}^{t} \dd{t'} \, \frac{\kappa_{\mu,\mathrm{c}}}{\sqrt{2\pi}}a(t') e^{ik \qty(t'-t)}  - i\int_{t_{\mathrm{in}}}^{t} \dd{t'} \, \frac{\kappa_{\mu,\mathrm{e}}}{\sqrt{2\pi}}\sigma_{-}(t')  e^{ik \qty(t'-t)}
		\\
		& \quad  - i\frac{\sum_{\nu \setminus \mu}V_{\mu \nu}}{\sqrt{2\pi}}\int_{t_{\mathrm{in}}}^{t} \dd{t'} \, \Phi_{\nu}(t') e^{ik \qty(t'-t)}. 
	\end{split}
\end{align}
We integrate over all $ k $ and divide through by $ \sqrt{2\pi} $
\begin{align}
	\begin{split}
		\Phi_{\mu}(t) -  b_{\mathrm{in},\mu}(t)  &= -i  \frac{\kappa_{\mu,\mathrm{c}}}{2}a(t)  - i \frac{\kappa_{\mu,\mathrm{e}}}{2}\sigma_{-}(t)   - i\frac{\sum_{\nu \setminus \mu}V_{\mu \nu}\Phi_{\nu}(t)}{2} 
	\end{split}
\end{align}
The factor of $ 1/2 $ comes from integrating over "half" the delta function. In matrix form, this becomes
\begin{align}
	\qty(\mathds{1}+\frac{i}{2}\mathbf{V})\boldsymbol{\Phi} &= \mathbf{b}_{\mathrm{in}}(t) -\frac{i}{2}\boldsymbol{\kappa}\mathbf{c}, \qquad \boldsymbol{\kappa} = \mqty(\boldsymbol{\kappa}_{\mathrm{c}} & \boldsymbol{\kappa}_{\mathrm{e}}). \label{eq:app_inOut_phi_in}
\end{align}
We note, that had we formally integrated from $ t $ to $ t_{\mathrm{out}} $ instead, we would have found
\begin{align}
	\qty(\mathds{1}-\frac{i}{2}\mathbf{V})\boldsymbol{\Phi} &= \mathbf{b}_{\mathrm{out}}(t) +\frac{i}{2}\boldsymbol{\kappa}\mathbf{c}. \label{eq:app_inOut_phi_out}
\end{align}
From Eq.~\eqref{eq:app_inOut_phi_in} and Eq.~\eqref{eq:app_inOut_phi_out} we find the first input-output relation
\begin{align}
	\mathbf{b}_{\mathrm{out}} = \mathbf{C}\mathbf{b}_{\mathrm{out}} + \mathbf{D}\mathbf{c},
\end{align}
with
\begin{align}
	\mathbf{C} = \qty(\mathds{1}-\frac{i}{2}\mathbf{V}) \qty(\mathds{1}+\frac{i}{2}\mathbf{V})^{-1},
\end{align}
and
\begin{align}
	\mathbf{D} = -i\qty(\mathds{1}+\frac{i}{2}\mathbf{V})^{-1} \boldsymbol{\kappa}.
\end{align}
To find the second input-output relation, we use Heisenberg equation of motion on the two system operators, $ a $ and $ \sigma_{-} $. This gives the following equations:
\begin{align}
	\partial_t a &= -i\omega_{\mathrm{c}}a -ig\sigma_{-} - i\mqty(1&0)\boldsymbol{\kappa}^{\dagger}  \qty(\mathds{1}+\frac{i}{2}\mathbf{V})^{-1} \mathbf{b}_{\mathrm{in}}(t) -\frac{1}{2}\mqty(1&0)\boldsymbol{\kappa}^{\dagger} \qty(\mathds{1}+\frac{i}{2}\mathbf{V})^{-1}\boldsymbol{\kappa}\mathbf{c},
\end{align}
and
\begin{align}
	\begin{split}
		\partial_t \sigma_{-} &=-i\omega_{\mathrm{e}}\sigma_{-} -ig^* a - i\mqty(0&1)\boldsymbol{\kappa}^{\dagger}\qty(\mathds{1}+\frac{i}{2}\mathbf{V})^{-1} \mathbf{b}_{\mathrm{in}} -\frac{1}{2}\mqty(0&1)\boldsymbol{\kappa}^{\dagger}\qty(\mathds{1}+\frac{i}{2}\mathbf{V})^{-1} \boldsymbol{\kappa}\mathbf{c}
		\\
		& \qquad  + 2ig^* \sigma_{+}\sigma_{-} a +2i\sigma_{+}\sigma_{-} \mqty(0&1)\boldsymbol{\kappa}^{\dagger}\qty(\mathds{1}+\frac{i}{2}\mathbf{V})^{-1}\mathbf{b}_{\mathrm{in}} + \sigma_{+}\sigma_{-} \mqty(0&1)\boldsymbol{\kappa}^{\dagger}\qty(\mathds{1}+\frac{i}{2}\mathbf{V})^{-1}\boldsymbol{\kappa}\mathbf{c}
	\end{split}
\end{align}
In matrix form this becomes
\begin{align}
	\dot{\mathbf{c}}(t) = \mathbf{A}\mathbf{c}(t)+\mathbf{B}\mathbf{b}_{\mathrm{in}}(t) + \mqty(0 \\ \hat{f}(t)), \qquad \mathbf{c}(t)=\mqty(a(t) \\ \sigma_{-}(t)), \label{eq:app_inOut_cDot}
\end{align}
with
\begin{align}
	\mathbf{A} = \mqty(-i\omega_{\mathrm{c}} & -ig \\ -ig^* & -i\omega_{\mathrm{e}}) -\frac{1}{2}\boldsymbol{\kappa}^{\dagger} \qty(\mathds{1}+\frac{i}{2}\mathbf{V})^{-1} \boldsymbol{\kappa},
\end{align}
\begin{align}
	\mathbf{B}&=-i \boldsymbol{\kappa}^{\dagger}\qty(\mathds{1}+\frac{i}{2}\mathbf{V})^{-1},
\end{align}
and
\begin{align}
	\hat{f}(t) = \qty(2ig^* + \mqty(0&1)\boldsymbol{\kappa}^{\dagger}\qty(\mathds{1}+\frac{i}{2}\mathbf{V})^{-1}\boldsymbol{\kappa}\mqty(1\\0) ) \sigma_{+}(t)\sigma_{-}(t) a(t) + 2i \mqty(0&1)\boldsymbol{\kappa}^{\dagger}\qty(\mathds{1}+\frac{i}{2}\mathbf{V})^{-1}\sigma_{+}(t)\sigma_{-}(t) \mathbf{b}_{\mathrm{in}}(t).
\end{align}
\subsection{$ S $-matrix calculations}
We first Fourier transform our input-output equation in Eq.~\eqref{eq:app_inOut_cDot} and rearrange terms:
\begin{align}
	\mathbf{c}(k) =i\qty(k\mathds{1}-i\mathbf{A})^{-1}\mathbf{B}\mathbf{b}_{\mathrm{in}}(k) + i\qty(k\mathds{1}-i\mathbf{A})^{-1} \mqty(0 \\ \tilde{f}(k)),
\end{align}
where
\begin{align}
	\begin{split}
		\tilde{f}(k) &= \qty(2ig^* + \mqty(0&1)\boldsymbol{\kappa}^{\dagger}\qty(\mathds{1}+\frac{i}{2}\mathbf{V})^{-1}\boldsymbol{\kappa}\mqty(1\\0) ) \int \frac{\dd{k'}}{\sqrt{2\pi}}\sigma_{+}(k') \sigma_{-} a (k+k')
		\\
		& \quad +2i \mqty(0&1)\boldsymbol{\kappa}^{\dagger}\qty(\mathds{1}+\frac{i}{2}\mathbf{V})^{-1}\int\frac{\dd{k'}\dd{k''}}{2\pi}\sigma_{+}(k')\sigma_{-}(k'') \mathbf{b}_{\mathrm{in}}(k+k'-k''). \label{eq:app_f_k}
	\end{split}
\end{align}
Using $ \langle \varnothing  | \hat{f} = 0 $, the single-photon $ S $-matrix (Eq.~\eqref{eq:S_pk_1ph}) can be expressed in terms of $ A, \, B, \, C, $ and $ D $ (see Eq.~{\eqref{eq:S1_JC}}):
\begin{align}
	\mathbf{S}^{(1)}_{p;k} = \qty[\mathbf{C} + i \mathbf{D}\qty(\mathds{1}p-i\mathbf{A})^{-1}\mathbf{B}]\delta(p-k) .
\end{align}
The two-photon $ S $-matrix is not so straight forward. We start from the definition and insert the input-output relations

\begin{align}
	S_{p_1p_2;k_1k_2}^{\mu_1\mu_2;\nu_1\nu_2} &= {}_{\mu_1 \mu_2}\braket{p_{1}p_{2}^-}{k_1k_2^+}_{\nu_1 \nu_2}= {}_{}\mel{\varnothing}{b_{ \mathrm{out},\mu_1}(p_1)b_{\mathrm{out},\mu_2}(p_2)}{k_1k_2^+}_{\nu_1 \nu_2}
	\\
	&=\sum_{\nu'} \int \dd{k'} {}_{}\mel{\varnothing}{b_{ \mathrm{out},\mu_1}(p_1)}{k'^+}_{\nu'}\,  {}_{\nu'}\mel{k'^+}{b_{\mathrm{out} , \mu_2}(p_2)}{k_1k_2^+}_{\nu_1 \nu_2}
	\\
	&= S_{p_1;k_1}^{\mu_1;\nu_1} S_{p_2;k_2}^{\mu_2;\nu_2} + S_{p_1;k_2}^{\mu_1;\nu_2}  S_{p_2;k_1}^{\mu_2;\nu_1} + iT_{p_1p_2;k_1k_2}^{\mu_1\mu_2;\nu_1\nu_2},
\end{align}
with
\begin{align}
	iT_{p_1p_2;k_1k_2}^{\mu_1\mu_2;\nu_1\nu_2} = \sum_{\nu'}S^{\mu_1;\nu'}(p_1) \, {}_{\nu'}\mel{p_1^-}{i\frac{iD_{\mathrm{c},\mu_2} A_{\mathrm{ce}}\hat{f}(p_2) + D_{\mathrm{e},\mu_2} \qty(p_2-iA_{\mathrm{cc}})\hat{f}(p_2) }{\qty(p_2-iA_{\mathrm{cc}})\qty(p_2-iA_{\mathrm{ee}})}}{k_1k_2^+}_{\nu_1 \nu_2}. \label{eq:app_BS_contr}
\end{align}

So far, these calculations have been completely general. In the two subsections below, we determine the matrix coefficients and $ S $-matrices in two special cases, first in case of a TLE coupled to a waveguide with a PTE, and second in case of a JC system coupled to a waveguide with a PTE.

\subsection{$ S $-matrix for a TLE}\label{app:smatrix_TLE}
For a TLE only, i.e. $ g=0 $ and $ \kappa_{\mathrm{c},\mu}=0 $ for all $ \mu $, Eq.~\eqref{eq:app_f_k} simplifies to
\begin{align}
	\begin{split}
		\tilde{f}(k) &= 2i \boldsymbol{\kappa}_{\mathrm{e}}^{\dagger}\qty(\mathds{1}+\frac{i}{2}\mathbf{V})^{-1}\int\frac{\dd{k'}\dd{k''}}{2\pi}\sigma_{+}(k')\sigma_{-}(k'') \mathbf{b}_{\mathrm{in}}(k+k'-k''). 
	\end{split}
\end{align}

The term for the bound state contribution (Eq.~\eqref{eq:app_BS_contr}) becomes
\begin{align}
	iT_{p_1p_2;k_1k_2}^{\mu_1\mu_2;\nu_1\nu_2} = \sum_{\nu'}S^{\mu_1;\nu'}(p_1) \, {}_{\nu'}\mel{p_1^-}{\frac{i D_{\mathrm{e},\mu_2} \hat{f}(p_2) }{\qty(p_2-iA_{\mathrm{ee}})}}{k_1k_2^+}_{\nu_1 \nu_2} .
\end{align}
We insert $ \tilde{f}(k) $ and find
\begin{align}
	iT_{p_1p_2;k_1k_2}^{\mu_1\mu_2;\nu_1\nu_2} =\frac{-i}{\pi}D_{\mathrm{e},\mu_2} \sum_{\nu'}S^{\mu_1;\nu'}(p_1) \mathcal{G}_{\nu'}^*(p_1) \mathcal{G}_{\nu_2}(p_2) \qty[\mathcal{G}_{\nu_1}(k_1)+\mathcal{G}_{\nu_1}(k_2)] \delta(p_1+p_2-k_1-k_2), \label{eq:app_iT2_TLE}
\end{align}
where we have defined
\begin{align}
	\mathcal{G}_{\nu}(k) &= \frac{-i\qty{\boldsymbol{\kappa}_{\mathrm{e}}^{\dagger}\qty(\mathds{1}+\frac{i}{2}\mathbf{V})^{-1}}_{\nu}}{k-iA_{\mathrm{ee}}}
\end{align}

For a Fano-waveguide symmetrically coupled to a lossless TLE, the matrices and matrix elements are
\begin{subequations}
	\begin{align}
		\mathbf{V} &= \mqty(0 & V  \\ V & 0 ),
		\\
		\boldsymbol{\kappa}_{\mathrm{e}} &= \mqty(\sqrt{\Gamma/2} \\ \sqrt{\Gamma/2}),
		\\
		A_{\mathrm{ee}} &= -i\omega_{\mathrm{e}} - \frac{\tilde{\Gamma}}{2},
		\\
		\mathbf{B}_{\mathrm{e}} &= -i\mqty(\frac{\tilde{\Gamma}}{\sqrt{\Gamma}} & \frac{\tilde{\Gamma}}{\sqrt{\Gamma}} ),
		\\
		\mathbf{D}_{\mathrm{e}} &= -i\mqty(\frac{\tilde{\Gamma}}{\sqrt{\Gamma}} \\ \frac{\tilde{\Gamma}}{\sqrt{\Gamma}} ).
	\end{align}
\end{subequations}

With this, Eq.~\eqref{eq:app_iT2_TLE} simplifies to
\begin{align}
	iT_{p_1p_2;k_1k_2}^{\mu_1\mu_2;\nu_1\nu_2} = \frac{1}{\pi}\frac{\tilde{\Gamma}}{\sqrt{2\Gamma}}\mathcal{G}(p_1)\mathcal{G}(p_2)\qty[\mathcal{G}(k_1)+\mathcal{G}(k_2)] \delta(p_1+p_2-k_1-k_2), 
\end{align}
with
\begin{align}
	\mathcal{G}(k) = \frac{\frac{\tilde{\Gamma}}{\sqrt{2\Gamma}}}{k-\omega_{\mathrm{e}}+i\frac{\tilde{\Gamma}}{2} }.
\end{align}

\subsection{$ S $-matrix for a JC system}\label{app:smatrix_JC}
Here, we consider the specific case of a Fano-waveguide symmetrically coupled to a lossless JC system. In this case, the emitter only couples to the waveguide via the cavity and not directly. We therefore take $ D_{\mathrm{e},\mu} $ to be zero for all $ \mu $ and the term for the bound state contribution (Eq.~\eqref{eq:app_BS_contr}) becomes
\begin{align}
	iT_{p_1p_2;k_1k_2}^{\mu_1\mu_2;\nu_1\nu_2} = \sum_{\nu'}S^{\mu_1;\nu'}(p_1) \, {}_{\nu'}\mel{p_1^-}{i\frac{iD_{\mathrm{c},\mu_2} A_{\mathrm{ce}}\hat{f}(p_2)  }{\qty(p_2-iA_{\mathrm{cc}})\qty(p_2-iA_{\mathrm{ee}})}}{k_1k_2^+}_{\nu_1 \nu_2} ,
\end{align}
where
\begin{align}
	\begin{split}
		\tilde{f}(k) &= 2ig^* \int \frac{\dd{k'}}{\sqrt{2\pi}}\sigma_{+}(k') \sigma_{-} a (k+k').
	\end{split}
\end{align}
Inserting this and assuming $ g=g^* $, we find
\begin{align}
	iT_{p_1p_2;k_1k_2}^{\mu_1\mu_2;\nu_1\nu_2} =\frac{2ig}{\sqrt{2\pi}} \sum_{\nu'}S^{\mu_1;\nu'}(p_1) \mathcal{G}_{\mathrm{e}}^*(p_1) \mathcal{G}_{\mathrm{e}}(p_2) \mel{\varnothing}{\sigma_{-} a (p_1+p_2)}{k_1 k_2^+}_{\nu_1 \nu_2} , \label{eq:app_iT2_JC_1}
\end{align}
where we have defined
\begin{align}
	\mathcal{G}_{\mathrm{e}}(k) &= \frac{\frac{\tilde{\Gamma}/2}{\sqrt{\Gamma/2}} g}{\qty[k- \omega_{\mathrm{c}} +i \frac{\tilde{\Gamma} }{2}]\qty[k -\omega_{\mathrm{e}} ] - g^2}     .
\end{align}
We also define
\begin{align}
	\mathcal{G}_{\mathrm{c}}(k) = \frac{\frac{\tilde{\Gamma}/2}{\sqrt{\Gamma/2}} \qty(k-\omega_{\mathrm{e}}) }{\qty[k- \omega_{\mathrm{c}} +i \frac{\tilde{\Gamma} }{2}]\qty[k -\omega_{\mathrm{e}} ] - g^2} .
\end{align}
Note that we here consider $ \sigma_{-} a (k) $ to be a single combined operator. In order to determine the matrix element $ \langle\varnothing | \sigma_{-} a (p_1+p_2) | k_1 k_2^+ \rangle_{\nu_1 \nu_2} $, we need a closed set of equations extending beyond the input-output equations. We find

\begin{align}
	\partial_t {\sigma_{-} a(t)} = \qty(-i\omega_{\mathrm{e}}\sigma_{-} a(t) -ig aa(t)  + f(t) a(t)) + \qty(\qty(-i\omega_{\mathrm{c}}-\frac{\tilde{\Gamma}}{2})\sigma_{-} a(t)  -i\frac{\tilde{\Gamma}/2}{\sqrt{\Gamma/2}} \sigma_{-}(t) \sum_{\mu=L,R}b_{\mathrm{in},\mu}(t)).
\end{align}

We see that an equation of motion for the combined operator $ aa $ is also needed:
\begin{align}
	\partial_t a^2(t) =  2\qty(-i\omega_{\mathrm{c}}-\frac{\tilde{\Gamma}}{2})aa(t) -2ig \sigma_{-} a(t) -2i\frac{\tilde{\Gamma}/2}{\sqrt{\Gamma/2}} a(t) \sum_{\mu=L,R}b_{\mathrm{in},\mu}(t). 
\end{align}

After Fourier transforming and rearranging terms, we find in matrix-vector form

\begin{align}
	\mqty(\sigma_{-} a(k) \\ aa(k)) =  - \frac{\tilde{\Gamma}/2}{\sqrt{\Gamma/2}}  \mqty(k - \qty(\omega_{\mathrm{c}} + \omega_{\mathrm{e}} ) +i\tilde{\Gamma}/2 & -g \\ -2g & k-2\omega_{\mathrm{c}} +i \tilde{\Gamma})^{-1} \int \frac{\dd{k'}}{\sqrt{2\pi}} \mqty(\sigma_{-}(k')-i\frac{\sqrt{\Gamma/2}}{\tilde{\Gamma}/2}\tilde{f a}(k) \\ 2a(k')  )\sum_{\mu} b_{\mathrm{in},\mu}(k+k').
\end{align}

We again use that $  \langle \varnothing  | \hat{f} = 0 $ (and therefore also $  \langle \varnothing  | \tilde{f a} = 0 $) and find

\begin{align}
	\mel{\varnothing}{\sigma_{-} a(p)}{k_1k_2^+}_{\nu_1 \nu_2}   &= - \frac{\frac{\tilde{\Gamma}/2}{\sqrt{\Gamma/2}}}{\sqrt{2\pi}} \frac{ \qty[p-2\omega_{\mathrm{c}} +i \tilde{\Gamma}]\qty(\mathcal{G}_\mathrm{e}(k_1)+\mathcal{G}_\mathrm{e}(k_2)) + 2g \qty(\mathcal{G}_\mathrm{c}(k_1)+\mathcal{G}_\mathrm{c}(k_2))  }{ \qty[p - \qty(\omega_{\mathrm{c}} + \omega_{\mathrm{e}} +i\tilde{\Gamma}/2)] \qty[p-2\omega_{\mathrm{c}} +i \tilde{\Gamma}] -2g^2 } \delta(p-k_1-k_2).
\end{align}

If we insert this into Eq.~\eqref{eq:app_iT2_JC_1} and simplify, we obtain the expression in Eq.~\eqref{eq:S_JC_nonlinearTerm}.
\end{widetext}
\section*{Acknowledgments}
K.B.J, J.I.-S., M.H., and J.M. acknowledge funding from the Danish Council for Independent Research (DFF-4181-00416). M.H. acknowledges support from the Villum Foundation. J.M. acknowledges support from the European Research Counsil (ERC) under the European Union Horizon 2020 Research and Innovation Programme (Grant no. 834410 Fano). Furthermore, the authors acknowledge helpful discussions with Yi Yu.
\newpage
\end{document}